\documentclass[twocolumn,aps,prd,preprintnumbers,superscriptaddress,nofootinbib,floatfix]{revtex4-1}
\usepackage{aas_macros}
\usepackage{here}
\usepackage[colorlinks = true,
            linkcolor = blue,
            urlcolor  = blue,
            citecolor = blue,
            anchorcolor = blue]{hyperref}
\usepackage{amsmath,amssymb,amsfonts,mathrsfs}
\usepackage[dvipdfmx]{graphicx}
\usepackage{hyperref}
\usepackage{color}
\usepackage{enumerate} 
\usepackage{bm}
\usepackage{array}
\usepackage{siunitx}
\usepackage{float}

\newcommand{\meyers}{\cite{Meyers:2020qrb}}

\definecolor{mygreen}{RGB}{0,115,0} 
 
\begin{document}
\title{Distinguishing a stochastic gravitational-wave signal from  correlated noise with joint parameter estimation: Fisher analysis for ground-based detectors}
\author{Yoshiaki Himemoto}
\affiliation{Department of Liberal Arts and Basic Sciences, College of Industrial Technology, Nihon University, Narashino, Chiba 275-8576, Japan}
\author{Atsushi Nishizawa}
\affiliation{Research Center for the Early Universe (RESCEU), School of Science, The University of Tokyo, Tokyo 113-0033, Japan}
\author{Atsushi Taruya}
\affiliation{Center for Gravitational Physics and Quantum Information, Yukawa Institute for Theoretical Physics, Kyoto University, Kyoto 606-8502, Japan}
\affiliation{
Kavli Institute for the Physics and Mathematics of the Universe, Todai Institutes for Advanced Study, the University of Tokyo, Kashiwa, Chiba 277-8583, Japan (Kavli IPMU, WPI)}
%
%
%
\date{\today}
\begin{abstract} 
Search sensitivity to a stochastic gravitational-wave background (SGWB) is enhanced by cross-correlating detector signals. However, one of the most serious concerns is the environmental noise correlated between detectors. The global electromagnetic fields on the Earth, known as Schumann resonances, produce the correlated noise through the instrumental magnetic couplings. In this paper, we study the detectability of a SGWB in the presence of the correlated magnetic noise, using the Fisher analysis based on the analytical model of the correlated magnetic noise. 
We find that there is no significant degeneracy between the SGWB and noise parameters. Marginalizing over the correlated noise parameters degrades the constraint on each SGWB parameter by a factor of $\sim2$ at most in the four-detector case, irrespective of the strength of the magnetic coupling.
We also confirm that the forecast results are robust against the variation of correlated noise parameters and can vary up to $40\%$ in the realistic range of the coupling parameters for the second-generation detectors. However, ignoring the correlated noise in parameter estimation generally leads to a biased constraint on the SGWB parameters. If the coupling strength is twice as large as expected, this could result in a serious bias.
\end{abstract}

\preprint{YITP-22-164}
\maketitle

\section{Introduction}
The direct detections of gravitational waves (GWs) by laser interferometers, LIGO (Hanford and Livingston)~\cite{LIGO1, LIGO2} and Virgo~\cite{Virgo} has opened up a new window to probe the Universe. Since its first discovery~\cite{GW150914:detection}, the number of GW events has increased dramatically ~\cite{LIGOScientific:2018mvr,LIGOScientific:2020ibl,LIGOScientific:2021djp}. The event rate of compact binary coalescences suggests that the expected number of such events over the cosmological scales would be too large to be individually resolved. A random superposition of GWs from such unresolved sources forms a stochastic GW background (SGWB) and would be detectable in the near future~\cite{KAGRA:2021kbb}.

To improve the sensitivity to a SGWB, a cross-correlation between detector signals has to be taken. However, one of the most serious concerns is the environmental noise correlated between detectors. It has been pointed out in Refs.~\cite{1992PhRvD..46.5250C, 1999PhRvD..59j2001A} that the (stationary) global electromagnetic fields on the Earth, known as Schumann resonances~\cite{1952ZNatA...7..149S, 1952ZNatA...7..250S}, produce the correlated noise through the instrumental magnetic couplings and give an impact on the searches for a SGWB. There are thus several experimental and theoretical studies to estimate the impact of correlated noise and try to mitigate its impact~\cite{2013PhRvD..87l3009T, 2014PhRvD..90b3013T, Coughlin:2016vor, Coughlin:2018tjc,  2017PhRvD..96b2004H, 2019PhRvD.100h2001H} 
(see also Refs.~\cite{Kowalska-Leszczynska:2016low, Washimi:2021ogz, Janssens:2022tdj} for short-duration transients).
While the recent study suggests that the correlated magnetic noise budget is still negligible with the current detector sensitivities~\cite{KAGRA:2021kbb}, but would be significant for 
LIGO, Virgo and KAGRA~\cite{KAGRA_2021PTEP} with the design sensitivity, and the future ground-based detectors~\cite{Janssens:2021cta}.

For the mitigation of the effect of the correlated magnetic noise, the analytical model has been developed~\cite{2017PhRvD..96b2004H, 2019PhRvD.100h2001H}. The model reproduces the major trend of the measured global magnetic correlation between the GW detectors and helps estimate the impact of the correlated noise on the detection of a SGWB. On the other hand, the analytical model is useful for forecasting the possibility to separate the correlated noise and to measure the parameters of a SGWB in the presence of the correlated magnetic noise. The methods to separate multiple components in a correlation signal have been developed in the cases of the mixture of cosmological and astrophysical GWs~\cite{Parida:2015fma, Martinovic:2020hru, Poletti:2021ytu} and the mixture of extra polarizations with the tensor modes in general relativity~\cite{Seto:2008sr, Nishizawa:2009bf}. For the component separation, the former method focuses on the difference of the spectral shapes of SGWBs and the latter utilizes the difference of the detector responses to each component.

Recently Meyers {\it et al.}~\cite{Meyers:2020qrb} evaluated the detectability of a SGWB in the presence of correlated magnetic noise with LIGO and Virgo in the framework of the Bayesian inference. Taking the coupling to the correlated magnetic noise into account in their parameter estimation, they found that the parameters of a SGWB spectrum can be determined well even when a strong correlated noise exists. However, the Bayesian framework is computationally expensive, and the parameter space of a SGWB and correlated magnetic noise has not been fully investigated. Furthermore, there is 1 more degree of freedom to characterize the correlated magnetic noise that has not been considered in Ref.~\cite{Meyers:2020qrb}. As Refs.~\cite{2017PhRvD..96b2004H, 2019PhRvD.100h2001H} advocated, the coupling to the magnetic field crucially depends on the direction to which the detector responds, and can significantly change the frequency dependence of the correlated magnetic noise. For these reasons, it is still unclear to what extent the correlated magnetic noise can affect the parameter estimation results. If we could not properly describe the correlated magnetic noise in the joint parameter estimation, the parameters of a SGWB can be biased. Such an impact has to be also studied quantitatively.   

In this paper, we discuss these issues based on the Fisher matrix formalism. 
In particular, based on the analytical model developed by Ref.~\cite{2017PhRvD..96b2004H, 2019PhRvD.100h2001H}, we shall study the detectability of a SGWB in the presence of correlated magnetic noise. Although the Fisher matrix analysis is an approximate treatment of the Bayesian statistical analysis and may sometimes give a wrong answer, it is expected to work well and can give an accurate estimation if the following conditions are met: (i) the likelihood function is nearly Gaussian with respect to the noise, (ii) the prior information is not important, and (iii) the nonlinearity in the model parameter space is negligible. 
In Appendix \ref{appendix:comparison_Myers_etal}, we make a comparison between our Fisher matrix analysis and the full Bayesian inference made by Ref.~\cite{Meyers:2020qrb}, and show that it is indeed the case, and our analysis produces results quantitatively consistent with their analysis.
Despite several cautious remarks on the use of the Fisher matrix analysis, one great advantage is that it is computationally inexpensive and enables easier exploration of the entire parameter space, in particular, of the correlated noise. Furthermore, it is physically more transparent and straightforward to interpret the results so that it would provide a practical way to mitigate the correlated magnetic noise in a realistic situation of GW observations.

This paper is organized as follows. In Sec.~\ref{sec:Schumann-resonance}, we briefly review the cross-correlation analysis and the analytical model of the correlated magnetic noise. In Sec.~\ref{sec:Fisher-matrix}, we explain the Fisher matrix analysis, introducing the formulas for the parameter estimation errors and biases. In Sec.~\ref{sec:results}, we present the results of the parameter estimation errors and biases and discuss the measurability of the parameters of a SGWB. We also discuss the dependence of the results on the variation of the magnetic noise parameters. Finally, Sec.\ref{sec:conclusion} is devoted to the summary of our findings and conclusions.

\section{Schumann resonance magnetic field }
\label{sec:Schumann-resonance}

In this section, we begin with briefly reviewing the standard cross-correlation analysis for detecting a SGWB in Sec.~\ref{subsec:cross_correlation}. We then consider in Sec.~\ref{subsec:magnetic_model} the correlated magnetic noise, and introduce an analytical model for the correlated noise presented in Refs.~\cite{2017PhRvD..96b2004H, 2019PhRvD.100h2001H}. The analytical model describes the global magnetic field and the coherence of the field between detectors, originating from Schumann resonances.
In Sec.~\ref{subsec:setup}, for the quantitative analysis of the impact of correlated noise on the parameter estimation of a SGWB in the later session, we summarize our baseline model inspired by Ref.~\meyers.

\subsection{Cross-correlation analysis}
\label{subsec:cross_correlation}

Let us first denote the time-series output data at $I$th detector by $s_{I}(t)$, which is composed of a
SGWB signal $h_{I}(t)$ and a noise $n_{I}(t)$:
\begin{align}
 s_{I}(t)=h_{I}(t)+n_{I}(t)\,.
\label{eq:output_data}
\end{align}
The SGWB signal is in most cases
considered to be very weak and random in nature. 
It is difficult to distinguish between a GW signal and a random instrumental noise 
from a single detector alone.  
Thus it is essential to detect it with the cross-correlation analysis by combining 
the output data from multiple detectors. 

Given the $I$th and $J$th output data during the observation time of $T_{\rm obs}$, we define the cross-correlation statistic $S$ by
\begin{align}
 S=\int_{-T_{\rm obs}/2}^{T_{\rm obs}/2}dt \int_{-T_{\rm obs}/2}^{T_{\rm obs}/2}dt'\, s_{I}(t)s_{J}(t') Q(t-t')\,.
\label{eq:correlation_signal}
\end{align}
Here the 
$Q$ is the filter function introduced to improve the detectability of the SGWB. 
If noises between two detectors are statistically uncorrelated, only a stochastic GW signal remains nonvanishing.
However, in the situation in which each detector is coupled to the global disturbances, these noises are correlated, leading to a nonzero cross-correlation statistic. Hence, it can be a contaminant for the detection of SGWBs. 

To consider the search for SGWBs in the presence of correlated noise, particularly arising from the Schumann resonances,
let us divide the noise $n_{I}$ in Eq.~(\ref{eq:output_data}) into two pieces: 
\begin{align}
 n_{I}(t)=n_{I}^{\rm inst}(t)+n_{I}^{\rm mag}(t)\,,
\label{eq:noise}
\end{align}
where $n_{I}^{\rm inst}$ and $n_{I}^{\rm mag}$ represent the instrumental noise due to local disturbances, and the correlated noise due to global magnetic fields on the Earth, respectively.
The expectation value of the cross-correlation statistic, $\langle S\rangle$, is expressed as a sum of the two contributions, i.e., $\langle h_{I} h_{J}\rangle$ and  $\langle n_{I}^{\rm mag} n_{J}^{\rm mag}\rangle$.
Assuming that the support of the filter function in the time domain is short enough compared to the observation time, the quantity $\langle S\rangle$ is expressed in the Fourier domain as 
\begin{align}
 \langle S \rangle = \int_{0}^{\infty}df\, U_{I J}(f)\,\widetilde{Q}(f)\,,
\label{eq:S_G_Fourier}
\end{align}
where the function $\widetilde{Q}$ is the Fourier transform of the filter function, which will be determined later so as to maximize signal-to-noise ratio (SNR). The function $U_{IJ}$ is the power spectral density and is expressed as
\begin{align}
U_{IJ}(f) &= U_{IJ}^{\rm gw}(f) + U_{IJ}^{\rm mag}(f) \,.
\label{eq:spectral-density}
\end{align}
The first term $U_{IJ}^{\rm gw}$ represents the spectral density
of the SGWB and is given by (e.g., Ref.~\cite{1999PhRvD..59j2001A})
\begin{equation}
U_{IJ}^{\rm gw}(f)= \frac{3H_0^2}{10\pi^2} \frac{\gamma_{IJ}^{\rm gw}(f) \Omega_{\rm gw} (f)}{f^3} \,,
\label{eq:spectral-gwb}
\end{equation}
where $H_0$ is the present Hubble parameter, and $\gamma_{I J}^{\rm gw}$ is the overlap reduction function which represents the coherence of the GW signals between the $I$th and $J$th detectors.
 The quantity $\Omega_{\rm gw}$ is the logarithmic energy density of a SGWB normalized by the critical density of the Universe, which is frequently parametrized in a power-law form as
\begin{equation}
\Omega_{\rm gw} (f)= \Omega_{{\rm gw},0} \left(\frac{f}{25\,{\rm Hz}}\right)^{n_{\rm gw}}\,.
\label{eq:gw-energy}
\end{equation}
In Eq.~(\ref{eq:spectral-density}), 
the power spectral density $U_{IJ}^{\rm mag}$ appears nonvanishing in the presence of the correlated magnetic noise, i.e., $\langle n_{I}^{\rm mag} n_{J}^{\rm mag}\rangle$. In the present paper, we will investigate its impact on the detection of SGWBs based on the analytical model developed in Refs.~\cite{2017PhRvD..96b2004H, 2019PhRvD.100h2001H}. In the next subsection, we present a brief overview of the model and introduce several important parameters.

\subsection{Correlated magnetic noise}
\label{subsec:magnetic_model}

One major source of the correlated noise, which we focus on in this paper, is the global magnetic field of the Earth-ionosphere cavity, known as the Schumann resonances \cite{1999PhRvD..59j2001A,2013PhRvD..87l3009T, 2014PhRvD..90b3013T}.

Coupled with the mirror systems of laser interferometers, the Schumann resonances can induce the nonvanishing  noise (i.e., $U_{IJ}^{\rm mag}$) that is correlated between two separated detectors.
Since the coupling between the mirror system and the global magnetic field is considered to be small, we assume that the correlated noise for the $I$th detector $n_I^{\rm mag}$ in Eq.~(\ref{eq:noise}) is linearly proportional to the global magnetic field ${\bm B}$ at the position ${\bm x}_{I}$. The correlated noise can then be expressed  in terms of its Fourier counterpart $\widetilde{\bm B}$ as follows:
\begin{align}
\widetilde{n}_{I}^{\rm mag}(f)=r_{I}(f)\,\left[{\widehat{\bm X}}_{I}\,\cdot \widetilde{{\bm B}}(f,{\bm x}_{I})\right].
\label{eq:conv_noise}
\end{align}
Here, the frequency-dependent quantity $r_I(f)$ 
characterizes the strength of the coupling between the detector and the magnetic field, which we call the coupling function.

Following Refs.~\cite{2013PhRvD..87l3009T, 2014PhRvD..90b3013T,2017PhRvD..96b2004H,2019PhRvD.100h2001H,Meyers:2020qrb}, we parametrize its functional form as
\begin{align}
r_{I}(f)=\kappa_I  \times 10^{-23}\left(\frac{f}{10 \,{\rm Hz}}\right)^{-\beta_I} [{\rm pT}^{-1}] \,.
\label{eq:coupling}
\end{align}
We assume that the parameters $\kappa_I$ and $\beta_I$ are constant in time. 
The unit vector $\widehat{\bm  X}_I$ describes the directional dependence of the coupling with magnetic fields. Since the Schumann resonances are described by a random superposition of the transverse magnetic modes of the electromagnetic waves, no vertical component of the magnetic field ${\bm B}$ exists idealistically \cite{1998clel.book.....J}. Hence, the directional coupling with mirror systems described by the vector $\widehat{\bm  X}_I$ is solely characterized by a single parameter $\psi_I$, which we call the orientation angle. This is an important parameter but the previous work on parameter estimation \cite{Meyers:2020qrb} has missed it. With particular attention to this coupling,
we will study the impact of the correlated magnetic noise on a SGWB search.

Provided the expression at Eq.~(\ref{eq:conv_noise}),
the spectral density  $U _{IJ}^{\rm mag} $ for the correlated noise is expressed in terms of the quantities in the Fourier domain as~\cite{2013PhRvD..87l3009T, 2017PhRvD..96b2004H}
\begin{align}
U _{IJ}^{\rm mag}  =  r_{I}(f)\,r_{J}(f)\,M_{I J}(f),  
\label{eq:S_B_Fourier}
\end{align}
where the function $M_{IJ}(f)$ characterizes the strength of the magnetic field and the coherence of the response to a pair of two separated detectors, and is further broken down into 
\begin{align}
 M_{IJ}(f) =
 M(f)\,\displaystyle{ \gamma_{IJ}^{\rm mag}(f) },
\label{eq:mij}
\end{align}
where $M(f)$ represents the global magnetic field spectrum, and $\displaystyle{ \gamma_{IJ}^{\rm mag}(f) }$ describes the magnetic coherence for a pair of detectors, and it is restricted to the range, $-1\leq\gamma_{IJ}^{\rm mag}\leq1$. Note that 
the function $\gamma_{IJ}^{\rm mag}$ is analogous to the GW overlap reduction function, $\gamma_{IJ}^{\rm gw}$, and its frequency dependence varies with a geometrical configuration of a detector pair and the orientation angle $\psi_{I/J}$ of each detector.
The definitions and explicit expressions for these functions are presented in Appendix \ref{appendix:analytic_formula} (see also Ref.~\cite{2017PhRvD..96b2004H}). 
In the next subsection, 
we introduce the baseline model of the correlated magnetic noise
and specify fiducial parameters to characterize its amplitude and frequency dependences.

\begin{figure}[t]
\begin{center}
\includegraphics[width=8.5cm]{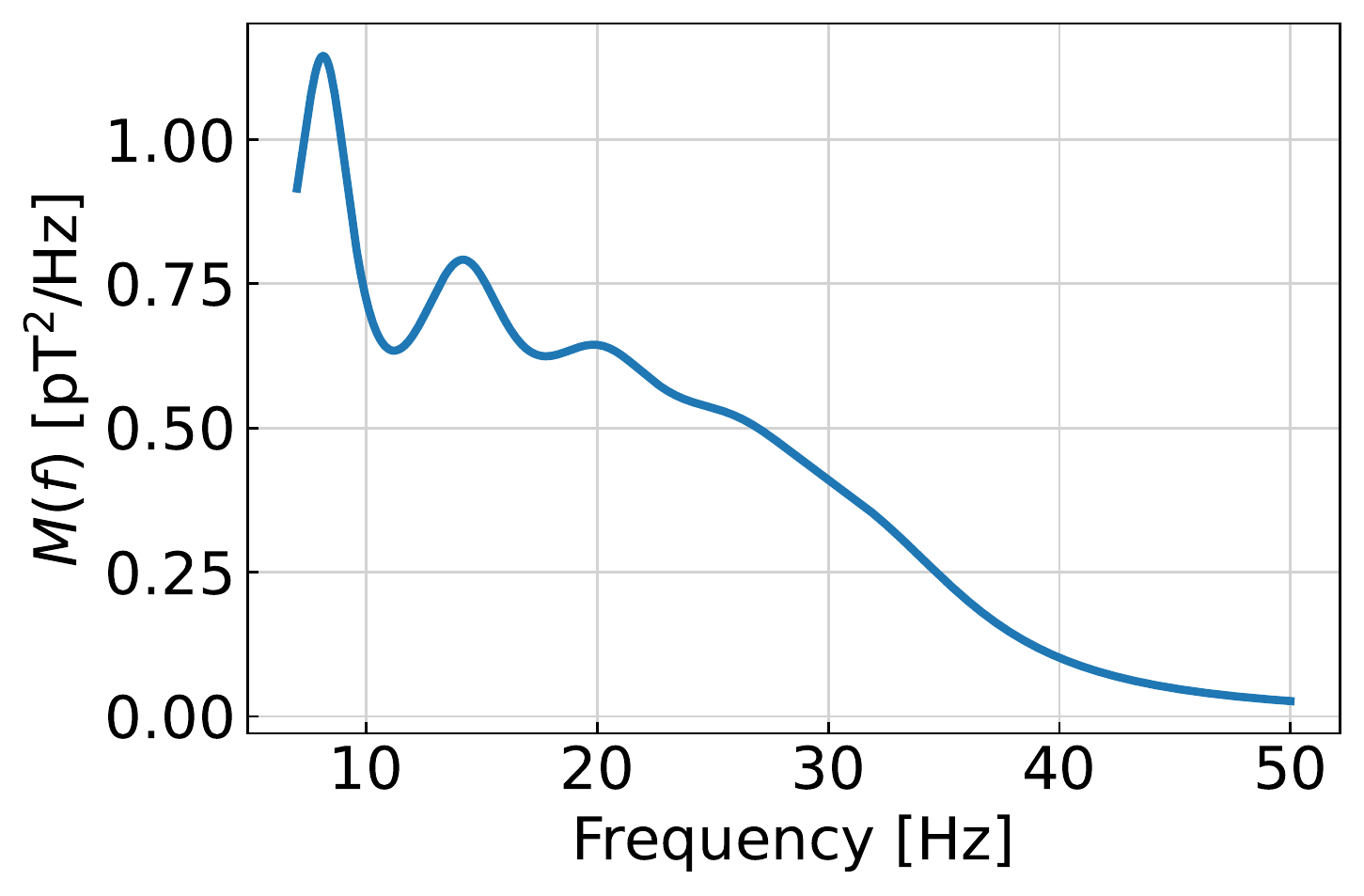}
\caption{Magnetic field spectrum, $M(f)$, as a function of frequency. The plotted result is obtained from the analytical expression and parameters summarized in Appendix \ref{appendix:analytic_formula}.}
\label{fig:magnetic-spectrum}
\end{center}
\end{figure}

\begin{figure*}[htb!]
\begin{center}
\includegraphics[width=17cm]{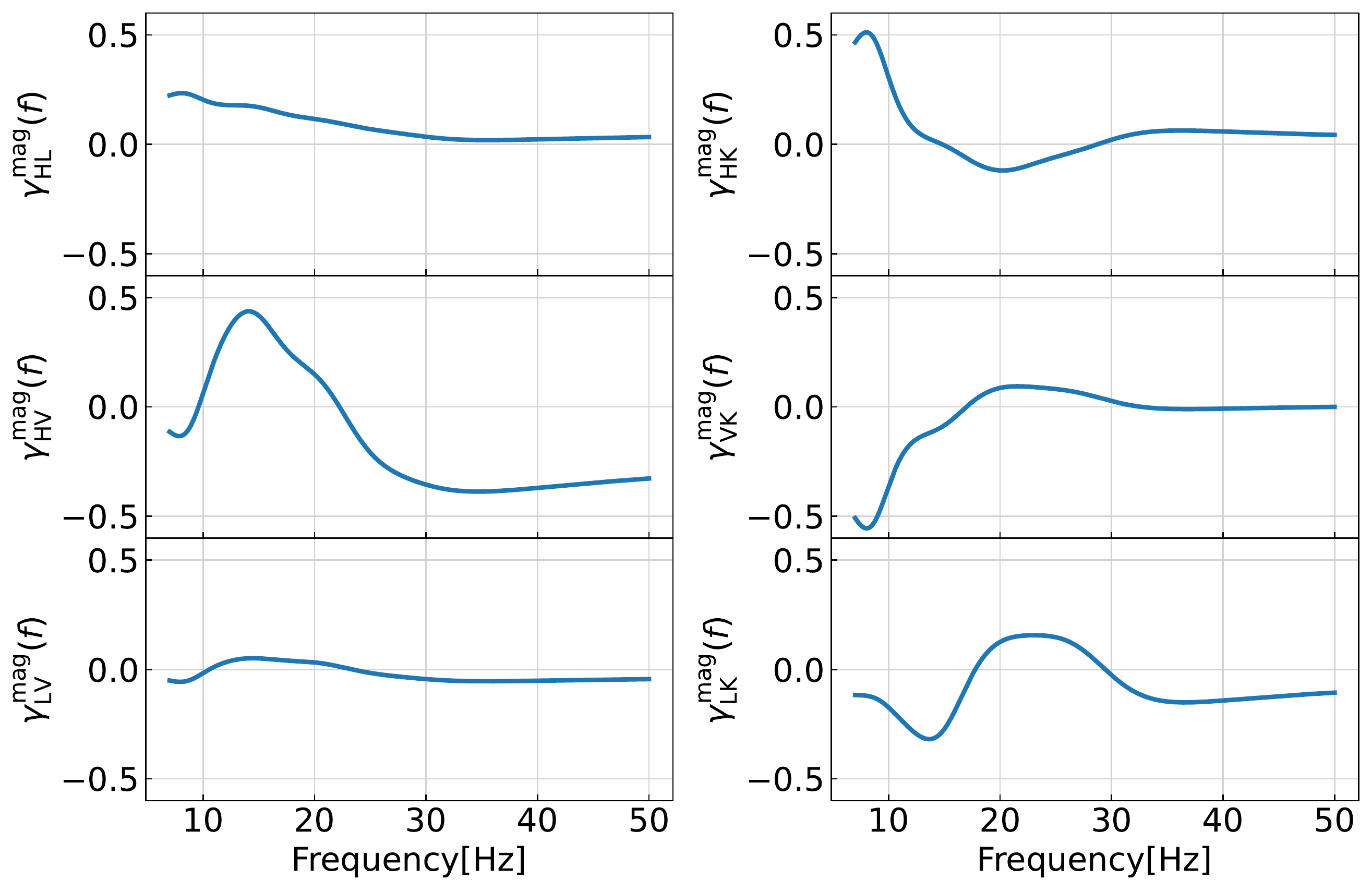}
\caption{Frequency dependence of magnetic coherence functions, $\gamma_{IJ}^{\rm mag}(f)$, for six representative pairs: LIGO Hanford and Livingston  (HL, upper left), LIGO Hanford and Virgo (HV, middle left), LIGO Livingston and Virgo (LV, lower left), 
Hanford and KAGRA (HK, upper right), Virgo and KAGRA (VK, middle right) and Livingston and KAGRA (LK, lower right). 
In plotting the results, the orientation angle $\psi_I$ for Hanford, Livingston and Virgo
are specified so as to reproduce the behaviors seen in Fig.~3 of Ref.~\cite{Meyers:2020qrb} in the frequency range of $10$--$30\,$Hz.
The orientation angle for KAGRA $\psi_{\rm K}$ is set to $\pi/2$ (see Table \ref{tab:magnetic-parameters}).
}
\label{fig:magnetic-coherence}
\end{center}
\end{figure*}

\subsection{Baseline models}
\label{subsec:setup}

In future observing runs, the joint parameter estimation considering both the 
SGWBs and correlated magnetic noise will become increasingly important, as shown in Refs.~\cite{KAGRA:2021kbb, Meyers:2020qrb}. 
Meyers {\it et al.}~\cite{Meyers:2020qrb} evaluated the parameter estimation errors of a SGWB in the presence of correlated magnetic noise based on Bayesian statistical inference. In this paper, we shall evaluate the statistical uncertainties of the SGWB and correlated-noise parameters, based on the Fisher matrix formalism. 
In what follows, we present the baseline model for the Fisher matrix analysis.

Let us first consider the correlated magnetic noise. For the function $M_{IJ}$, we set the parameters characterizing $M$ so as to reproduce Fig.~2 of Ref.~\cite{Meyers:2020qrb}.  In Appendix \ref{appendix:analytic_formula}, a specific choice of the parameter set is presented. Based on this, Fig.~\ref{fig:magnetic-spectrum} shows the function $M$. On the other hand, for the coherence function $\gamma_{IJ}^{\rm mag}$ in Eq.~(\ref{eq:mij}), we looked for the orientation angle $\psi_I$ of LIGO Hanford (H), Livingston (L) and Virgo (V) so as to reasonably reproduce the measured coherence by the magnetometers in Fig.~3 of Ref.~\cite{Meyers:2020qrb}. We did this particularly focusing on the frequencies of $10-30$ Hz, where the detectors have the best sensitivity to the SGWB. For KAGRA (K), since we do not have specific information on its magnetic coherence, we simply set $\psi_K$ to $\pi/2\simeq 1.57$. As we will show below, the choice of $\psi_K$ does not change much the results. Our setup of the parameter $\psi_I$, measured counterclockwise from the local East direction, is summarized in Table~\ref{tab:magnetic-parameters}. Using these parameters, the magnetic coherence functions are plotted in  Fig.~\ref{fig:magnetic-coherence} for all pairs of detectors, HLVK.

\begin{table}[tb]
\caption{Magnetic coupling parameters and orientation angles (in units of radians) used in the Fisher matrix analysis. }
\begin{center}
\begin{tabular}{|l|c|c|c|}
\hline
Detectors & $\kappa_I$ & $\beta_I$ & $\psi_I$ \\
\hline\hline
LIGO (Hanford) & 0.38 & 3.55 & 5.97 \\
LIGO (Livingston) & 0.35 & 4.61 & 0.64 \\
Virgo & 0.275 & 2.50 & 1.12 \\
KAGRA & 0.38 & 2.50 & 1.57 \\
\hline
\end{tabular}
\end{center}
\label{tab:magnetic-parameters}
\end{table}

\begin{figure}[t]
\begin{center}
\includegraphics[width=8.cm]{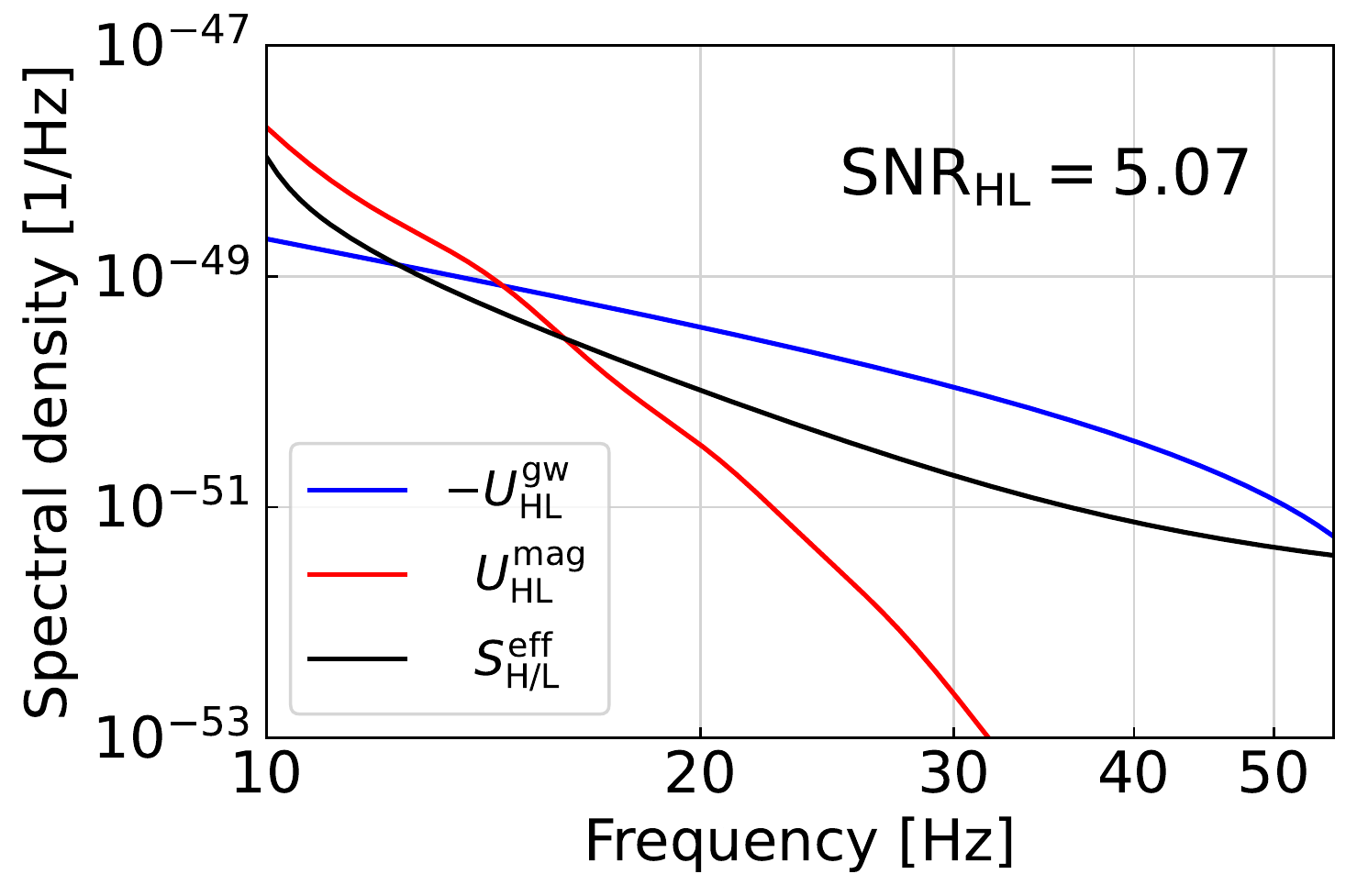}
\caption{
Power spectral densities of SGWB (blue) and correlated magnetic noise (red) for the LIGO HL pair, i.e., $U_{\rm HL}^{\rm gw}$ and $U_{\rm HL}^{\rm mag}$, 
which are compared with the effective spectral density of the instrumental noise $S_{H/L}^{\rm eff}$ defined in Eq.~(\ref{eq:eff_noise}) (black). Here, the function $U_{\rm HL}^{\rm gw}$ takes negative values and the plotted result is multiplied by $-1$.
In computing $U_{\rm HL}^{\rm gw}$, we adopt $\Omega_{\rm gw}(f)$ in Eq.~(\ref{eq:gw-energy}), with $(\Omega_{{\rm gw},0}, n_{\rm gw})=(3\times10^{-9}, 2/3)$
and the observation time $T_{\rm obs} =1$yr, which yields the signal-to-noise ratio of $\mathrm{SNR_{HL}}=5.07$. 
As for $U_{\rm HL}^{\rm mag}$, we adopt the coupling function in Eq.~(\ref{eq:coupling}) and the coherence function in Eq.~(\ref{eq:coherence}), 
adopting the coupling parameters in Table \ref{tab:magnetic-parameters}.}
\label{fig:spectral-densities}
\end{center}
\end{figure}

For the coupling function introduced at Eq.~(\ref{eq:coupling}), following Ref.~\meyers, we adopt the coupling parameters $(\kappa_{I}, \beta_{I})$ based on measurements made after the second observation run (O2) at LIGO Hanford and Livingston, and post-O2 measurements at Virgo. A set of the parameters above is called the realistic parameters in the present paper.
For KAGRA, the coupling strength $\kappa_I$ and slope $\beta_I$ were selected to be the worst (the largest and smallest, respectively) case among the three detectors.
They are also listed in Table \ref{tab:magnetic-parameters}. 

\begin{figure}[t]
\begin{center}
\includegraphics[width=8.5cm]{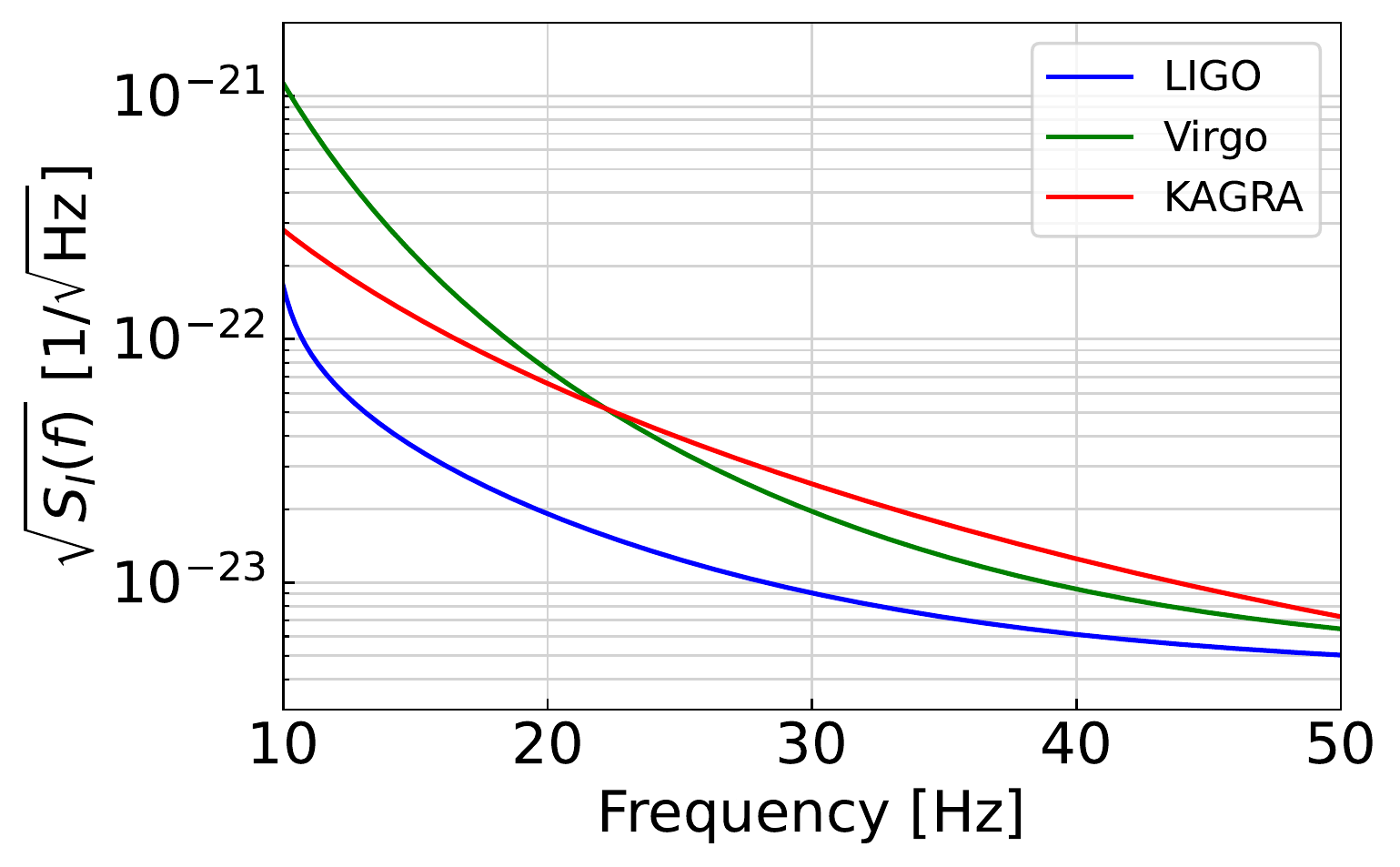}
\caption{Square root of the instrumental noise power spectral density, $\sqrt{S_{I}(f)}$, for second generation detectors, LIGO (blue), Virgo (green) and KAGRA (red). The plotted curves are for the design sensitivity, and we assume that the noise power spectral densities for LIGO Hanford and Livingston are identical. 
}
\label{fig:noise-curves}
\end{center}
\end{figure}

The main task of this paper is to evaluate the impact of correlated magnetic noise 
on the parameter estimation for SGWB, based on a Fisher matrix analysis.
Figure~\ref{fig:spectral-densities} illustrates the frequency dependences of the spectra of an astrophysical SGWB (see below), $U_{IJ}^{\rm gw}$ (blue), and correlated magnetic noise, $U_{IJ}^{\rm mag}$ (red), with our fiducial setup parameters (Table~\ref{tab:magnetic-parameters}).
It is well known that the size of the estimation errors of the SGWB depends on signal-to-noise ratio (SNR), defined as the ratio of the cross-correlation statistic to its standard deviation.
Considering the weak-signal limit, in which the variance of the cross-correlation statistic $S$, defined by $\sigma^{2} \equiv \langle S^{2} \rangle-\langle S \rangle^{2}$, is dominated by the detector's noise (i.e. $|h_I|\,,\, |n_I^{\rm mag}| \ll| n_I^{\rm inst}|$), and the optimal filter $Q$ to maximize the SNR for a SGWB in Eq.~(\ref{eq:spectral-gwb}), the square of SNR for the $I$th and $J$th detector pair is given by~\cite{1999PhRvD..59j2001A}
\begin{equation}
{\rm SNR}_{IJ}^2 = 2 T_{\rm obs} \int_0^{\infty} \frac{\{U_{IJ}^{\rm gw}(f) \}^2}{S_I(f) S_J(f)} df \;,
\label{eq:snr-gwb}
\end{equation}
where the function $S_{I}(f)$ is the instrumental noise power spectral density for the $I$th detector. 
In the following estimation, for  $S_{I}$, we use the table of numerical data for the design sensitivity of LIGO Hanford/Livingston \cite{Shoemaker:2014}  and the fitting form of the noise spectra for KAGRA and Virgo, given in Eqs.~(5) and (6) of Ref.~\cite{2012arXiv1202.4031M}, respectively (see  Fig.~\ref{fig:noise-curves}). 
We then define the effective power spectral density for the instrumental noise [see Eq.~(\ref{eq:snr-gwb})] as
\begin{equation}
S_I^{\rm eff}(f) = \frac{S_I(f)}{\sqrt{2T_{\rm obs} f}} \;.
\label{eq:eff_noise}
\end{equation}
The black line in Fig.~\ref{fig:spectral-densities} presents  the effective power spectral density
for the LIGO instrumental noise, $S_{\rm H/L }^{\rm eff}(f)$.

For the parametrization of the energy density of SGWB in Eq.~(\ref{eq:gw-energy}), we consider an astrophysical SGWB generated by compact binary coalescences, and set $\Omega_{{\rm gw},0}$ and $n_{\rm gw}$ to $3\times10^{-9}$ and $2/3$, respectively. The former 
comes from the current upper limit~\cite{KAGRA:2021kbb}. 
With these parameters of a SGWB, the ${\rm SNR_{HL}}$ is 5.07 for the observation time $T_{\rm obs} = 1$ year.
Note that the LIGO pair has dominantly a better sensitivity compared to other detector pairs, and the total SNR combining more than three detectors, defined by ${\rm SNR}_{\rm tot}\equiv \{\sum_{I,J}{\rm SNR}_{IJ}^2\}^{1/2}$, remains almost the same as the one for the HL pair, i.e., ${\rm SNR}_{\rm tot}=5.16$ for the three detector among H, L, and V, and $5.20$ for the four-detector case when further adding KAGRA.
Figure~\ref{fig:spectral-densities} shows that $U_{IJ}^{\rm mag}$ is dominant at low frequencies and $U_{IJ}^{\rm gw}$ becomes dominant above 
$15$ Hz. In general, such a difference of the spectral tilts is beneficial in separating a mixture of two components. In addition, the differences between $\gamma_{IJ}^{\rm gw}$ in Eq.~(\ref{eq:spectral-gwb}) and $\gamma_{IJ}^{\rm mag}$ in Eq.~(\ref{eq:mij}) also help us to distinguish a SGWB signal from correlated noise in Sec.~\ref{sec:results}. 

\section{Fisher matrix analysis}
\label{sec:Fisher-matrix}
In this section, based on the Fisher matrix formalism, we present the prescription to estimate the statistical errors and systematic biases of the parameters in a  mixture of a SGWB signal and correlated magnetic noise.

\subsection{Statistical errors}
\label{subsec:Fisher-error}

Given a correlation signal between $I$th and $J$th detectors, $U_{IJ}$ defined in Eq.~(\ref{eq:spectral-density}), the Fisher matrix is given by (e.g.,~Refs.~\cite{Seto:2005qy,Kuroyanagi_2018JCAP})
\begin{equation}
F_{ab} = 2\, T_{\rm obs} \sum_{(I,J)} \int_{f_{\rm min}}^{f_{\rm max}}
 \frac{\partial_{a} U_{IJ}(f)\, \partial_{b}
 U_{IJ}(f)}{S_I(f)S_J(f)} df \;,
\label{eq:Fisher-matrix} 
\end{equation}
where the model parameters in $U_{IJ}$ are
\begin{equation}
\theta_a=\{\Omega_{{\rm gw},0}, n_{\rm gw}, \kappa_I, \beta_I, \psi_I \}\;.
\end{equation}
The symbol $\partial_a$ stands for the derivative with respect to a parameter $\theta_a$. The sum is taken over all detector pairs, $(I,J)$. $T_{\rm obs}$ is the observation time. The lower cutoff of the frequency integral is set to $f_{\rm min}=10\,{\rm Hz}$, while the higher cutoff is set to $f_{\rm max}=200\,{\rm Hz}$ so that the integral converges.
For three detectors (HLV) and four detectors (HLVK), the number of the parameters is at maximum 11 and 14, respectively. 
Provided the Fisher matrix, the statistical error of a parameter marginalized over others, which we denote by $\delta\theta_a$, is estimated to be 
\begin{equation}
\delta \theta_{a}=\sqrt{(F^{-1})_{aa}},
\label{eq:Inverse-Fisher-matrix} 
\end{equation}
where the matrix $(F^{-1})_{ab}$ is the inverse of Fisher matrix. The errors are basically scaled with the inverse of SNR.

\subsection{Systematic biases}
\label{subsec:Fisher-bias}

Given the likelihood function, the Fisher matrix formalism also provides a simple way
to estimate the biases in the best-fit parameters caused by an incorrect assumption on a model. In the presence of a global magnetic field, what is likely to occur is modeling a correlation signal without knowing the presence of the magnetic field and misinterpreting the observational data as a correlation signal from a SGWB. This can potentially affect the best-fit parameters for the model in which both a SGWB and a magnetic field are assumed to exist.

Consider a correlation signal in the presence of a magnetic correlated noise:
\begin{align}
U_{IJ}^{\rm true} = U_{IJ}^{\rm gw}+U_{IJ}^{\rm mag} \;.
\end{align}
If the correlation signal is analyzed under the assumption that the signal is totally from a SGWB, that is, $U_{IJ} = U_{IJ}^{\rm gw}$, the systematic deviation is
\begin{align}
U_{IJ}^{\rm sys} = U_{IJ}- U_{IJ}^{\rm true} =-U_{IJ}^{\rm mag} \;.
\end{align}
By misinterpreting $U_{IJ}^{\rm sys}$ as a part of the SGWB signal, the best-fit parameters are biased from the true values $\theta_{a}^{\rm true}$ by (see, e.g.~Appendix of Ref.~\cite{Himemoto:2021ukb} for the derivation)
\begin{align}
\Delta \theta_{a}&=\sum_{b} \bigl(\mathcal{F}^{-1}\bigr)_{ab}\,s_b    
\label{eq:formula_systematic_bias}
\end{align}
with the matrix $\mathcal{F}_{ab}$ and the vector $s_a$, respectively, given by
\begin{align}
\mathcal{F}_{ab} &= 2 T_{\rm obs} \sum_{(I,J)} \int_{0}^{\infty} \frac{1}{S_I(f) S_J(f)} \left\{ \partial_a U_{IJ}^{\rm gw}(f) \partial_b U_{IJ}^{\rm gw}(f) \right. \nonumber \\ 
& \qquad \qquad \qquad \qquad \left. -U_{IJ}^{\rm mag}(f)\,\partial_a\partial_b U_{IJ}^{\rm gw}(f) \right\} \,df,
\label{eq:matrix_F}
\\
s_a&=2 T_{\rm obs} \sum_{(I,J)} \int_{0}^{\infty} \frac{U_{IJ}^{\rm mag}(f)\partial_aU_{IJ}^{\rm gw}(f) }{S_I(f) S_J(f)}\,df.
\label{eq:vector_s}
\end{align}
Note that the integrands in these expressions are evaluated at the fiducial (true) parameters. When the magnetic noise is negligible, that is, $|U_{IJ}^{\rm mag}|\ll|U_{IJ}^{\rm gw}|$, the matrix $\mathcal{F}_{ab}$ is reduced to the Fisher matrix for the SGWB model in the absence of the magnetic noise, $F_{ab}$. Then the size of the systematic bias $\Delta\theta_a$ is linearly proportional to the ratio of the magnetic noise amplitude to the GW amplitude, and hence roughly scales with the ratio of the SNRs.

\begin{figure}[t]
\begin{center}
\includegraphics[width=8.5cm]{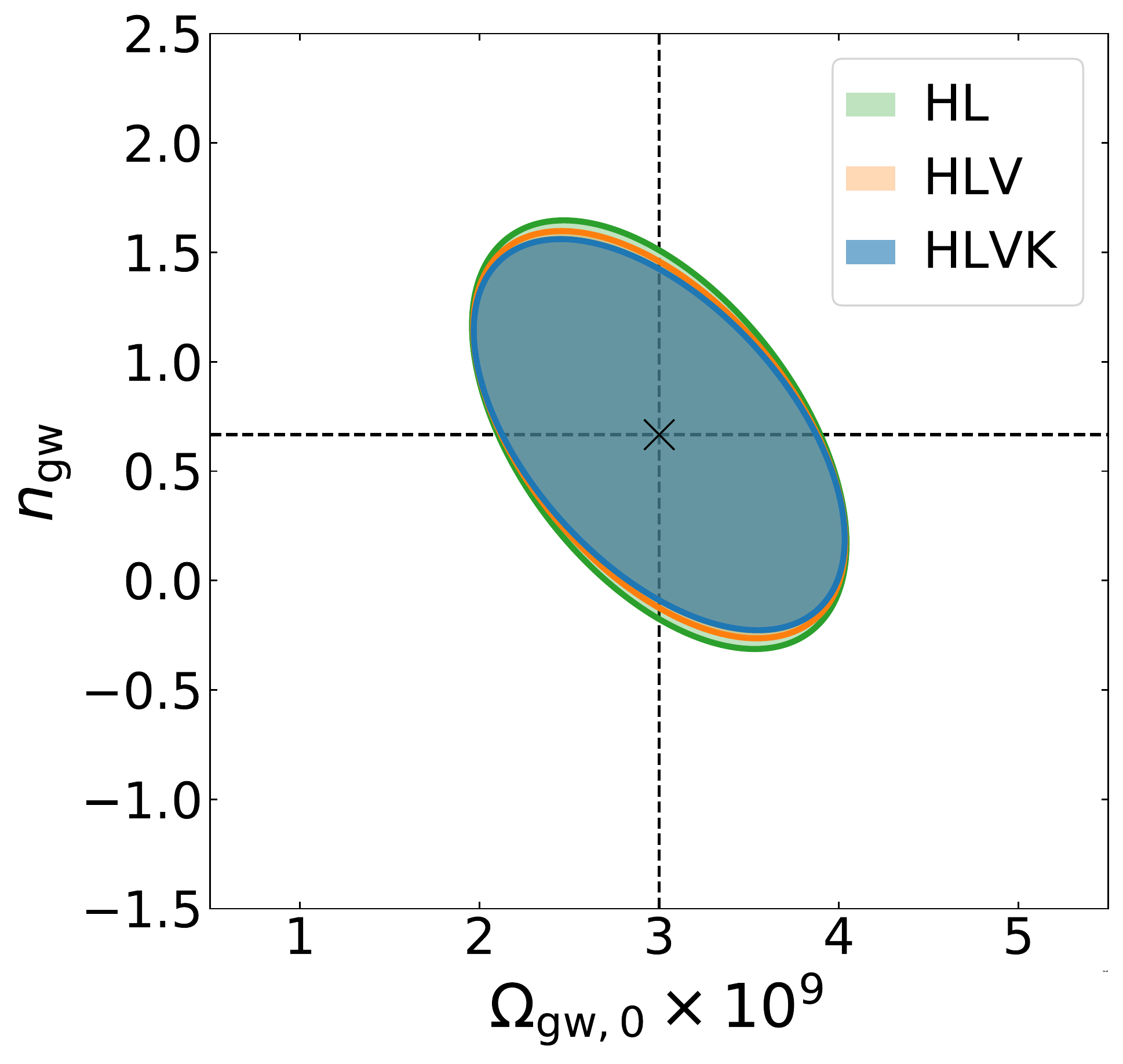}
\caption{Forecast constraints on SGWB parameters, $\Omega_{\mathrm{gw},0}$ and $n_{\rm gw}$ in the absence of correlated magnetic noise. Green, orange and blue shaded contours correspond to the expected 1$\sigma$ errors obtained from HL pair, three-detector network of HLV, and four-detector network of HLVK, respectively, assuming the observation time of $T_{\rm obs}=1$ yr and the design sensitivity for each detector. 
}
\label{fig:Error_Contour_wo_Mag}
\end{center}
\end{figure}
\begin{figure*}[t!]
\begin{center}
\includegraphics[width=8.5cm]{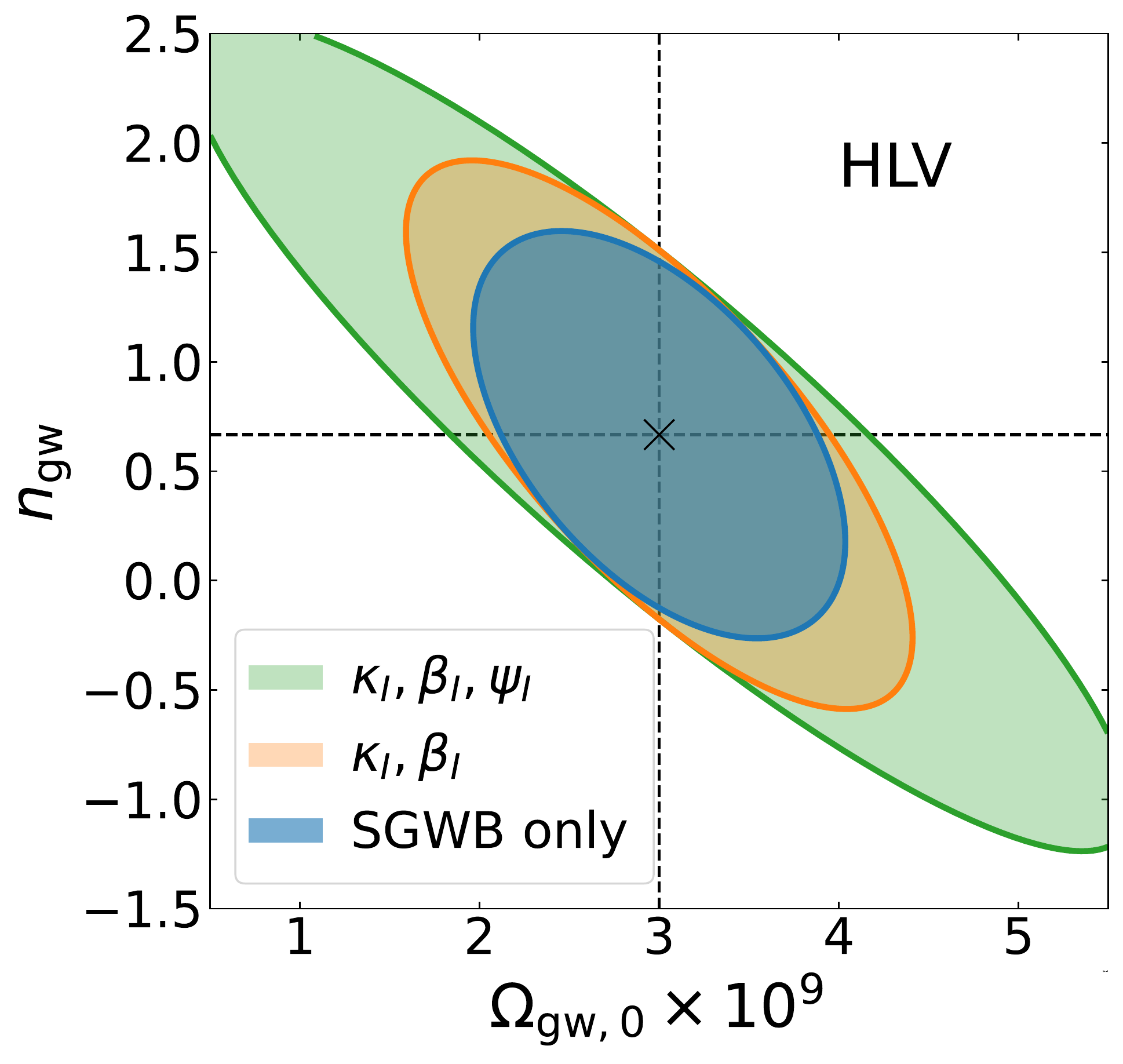}
\includegraphics[width=8.5cm]{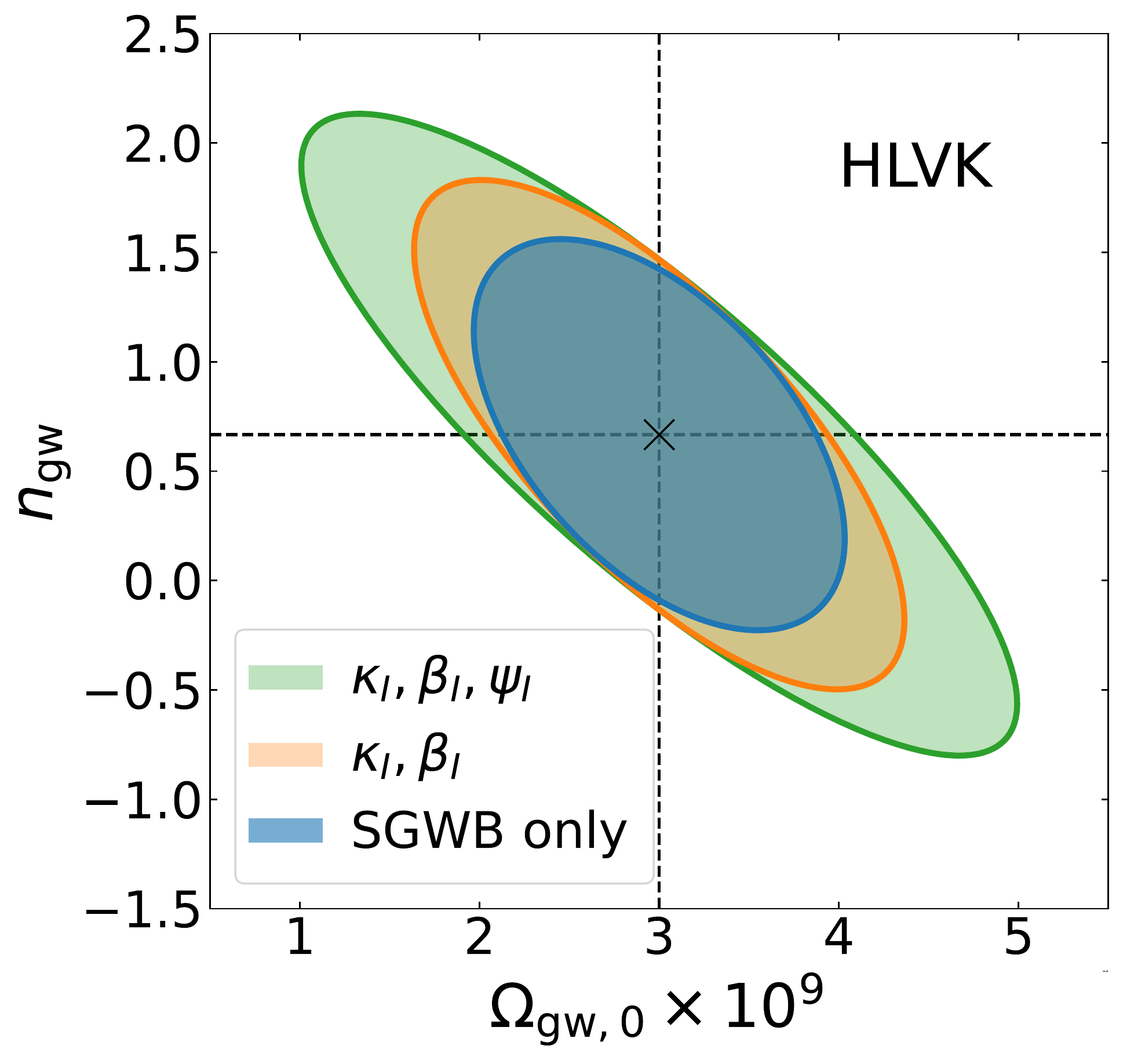}
\caption{Forecast constraints on $\Omega_{\mathrm{gw},0}$ and $n_{\mathrm{gw}}$ in the presence of correlated magnetic noise. Left and right panels are the results obtained from the three- and four-detector networks of HLV and HLVK, respectively. 
In each panel, green and orange shaded contours represent the $1\sigma$ statistical errors obtained from the joint parameter estimation of both the SGWB and correlated magnetic noise parameters. 
While the green contours are the case where both the coupling parameters $(\kappa_I, \,\beta_I)$ and the orientation angle $\psi_I$ are varied, the orange contours are the results when we consider only the coupling parameters, with the orientation angles kept fixed and excluded from the Fisher matrix. For reference, the blue shaded contours in the left and right panels are the results in the absence of correlated noise, identical to the orange and blue shaded contours in Fig. ~\ref{fig:Error_Contour_wo_Mag}, respectively.
}
\label{fig:Error_Contour_w_Mag}
\end{center}
\end{figure*}

\begin{figure*}[htb!]
\begin{center}
\includegraphics[width=18cm]{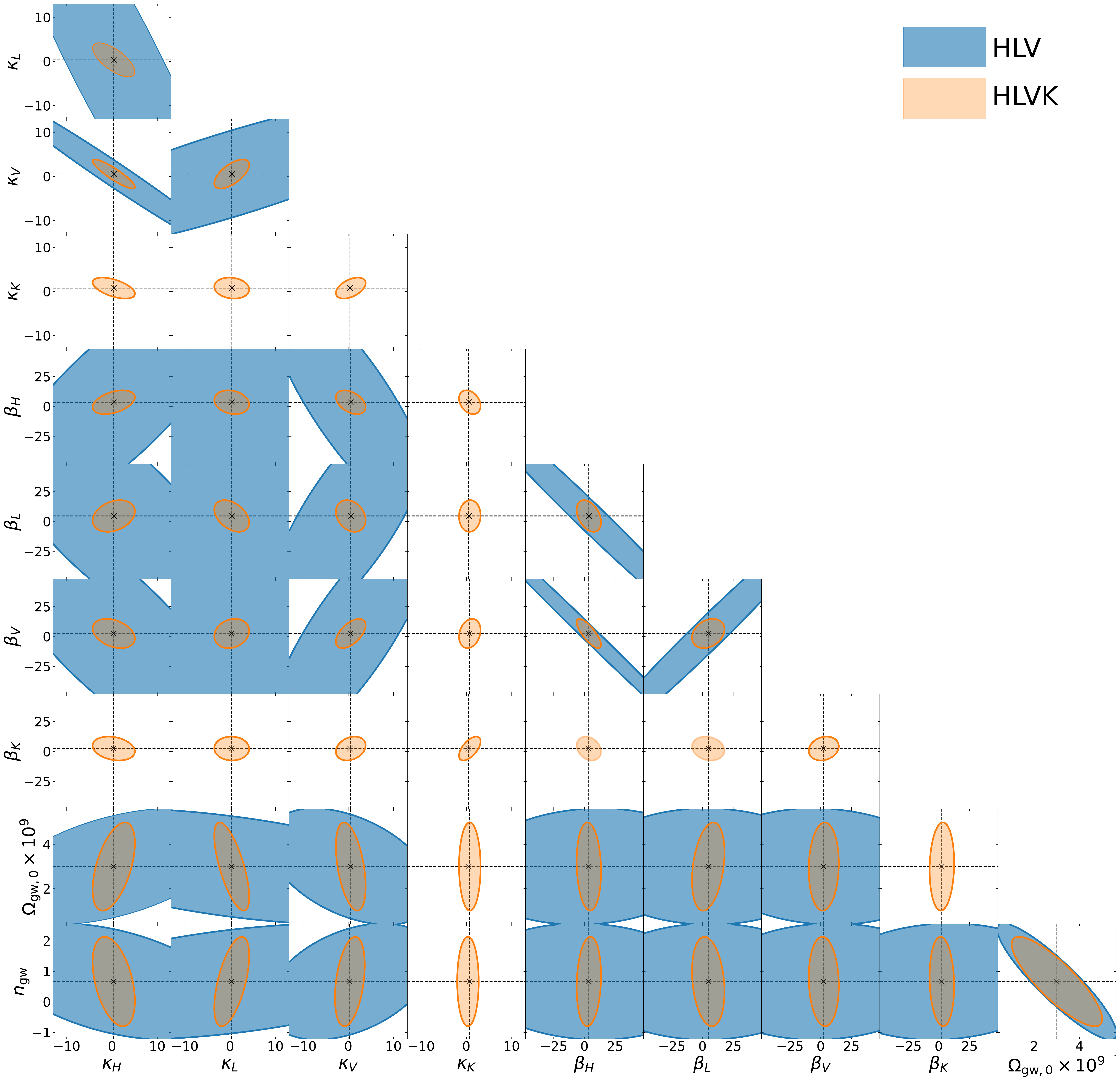}\\
\caption{Expected error contours on both the SGWB parameters and the coupling parameters of the correlated noise, with the orientation angle $\psi_{I}$ marginalized over. Blue and orange shaded contours represent the $1\sigma$ statistical error for the three-detector (HLV) and four-detector (HLVK) cases, respectively. 
Note that the blue and orange shaded contours in the bottom rightmost panel are identical respectively to the green shaded contours shown in the left and right panels of Fig.~\ref{fig:Error_Contour_w_Mag}.}
\label{fig:Contour_Meyer_design}
\end{center}
\end{figure*}

\section{Results}
\label{sec:results}

In what follows, setting the observation time $T_{\rm obs}$ to $1$ year and adopting 
the magnetic coupling parameters shown in Table \ref{tab:magnetic-parameters}, we present the results for the Fisher matrix analysis. In Sec.~\ref{subsec:parameter_estimation_w_correlated_noise}, 
we consider the joint parameter estimation taking the correlated noise parameters into account, and discuss how the constraints on the SGWB parameters are degraded in the presence of correlated magnetic noise. Also we discuss an impact of the network detection of a SGWB, and see how increasing the number of detectors helps tightly constrain the correlated noise parameters, hence leading to an improved constraint on a SGWB. In Sec.~\ref{subsec:systematic_biases}, we examine the case ignoring the correlated magnetic noise, and estimate how the SGWB parameters are biased from their fiducial values. Throughout the analysis, 
we fix the Hubble parameter $H_0$ to $70\,$km\,s$^{-1}$\,Mpc$^{-1}$, and adopt the baseline model for the fiducial values of the SGWB parameters, i.e.,~$(\Omega_{\rm gw,0},\,n_{\rm gw}) = (3 \times 10^{-9},\,2/3)$.

\subsection{Parameter estimation with correlated magnetic noise}
\label{subsec:parameter_estimation_w_correlated_noise}

In this subsection, taking the correlated magnetic noise into account, its impact on the parameter estimation of a SGWB is presented. To do so, we first show in Fig.~\ref{fig:Error_Contour_wo_Mag} the expected constraints on the SGWB in the absence of correlated magnetic noise. That is, we consider only the amplitude and spectral index of SGWB, $\Omega_{\rm gw,0}$ and $n_{\rm gw}$, and plot the parameter estimation errors in the cases with two detectors (HL), three detectors (HLV), and four detectors (HLVK). In general, as increasing the number of detectors to combine, the constraining power on the  SGWB is improved. However, the result in Fig.~\ref{fig:Error_Contour_wo_Mag} shows that the combination of detectors other than the LIGO pair (HL) does not help so much in improving the SGWB constraint. This is because the sensitivity of the LIGO detector pair is the best among others, because of the best noise spectral density at $f\lesssim50$Hz and the closest separation.

We next consider the correlated magnetic noise, taking the model parameters of the magnetic noise into account, and constrain the magnetic noise parameters simultaneously. Note that in this case, unless we consider more than three detectors, the Fisher matrix becomes singular and we cannot evaluate the inverse matrix. Hence, in Fig.~\ref{fig:Error_Contour_w_Mag}, the expected errors on $\Omega_{\rm gw,0}$ and $n_{\rm gw}$ are presented for the three-detector (left) and four-detector (right) cases, with the magnetic noise parameters marginalized over. In each panel, blue shaded contours are just a reference to the case without the correlated magnetic noise shown in Fig.~\ref{fig:Error_Contour_wo_Mag}. On the other hand, the contours depicted as orange and green colors are the results allowing the coupling parameters $(\kappa_I,\beta_I)$ and those with the orientation angle $(\kappa_I,\beta_I,\psi_I)$ to vary, respectively. That is, including the two parameters characterizing a SGWB, the number of free parameters in the Fisher matrix is 8 for the former and 11 for the latter in the case of three-detector case (left, HLV), and they are increased to 10 and 14 when we consider the four-detector case (right, HLVK). 

Clearly, the statistical errors are inflated as increasing the number of parameters. 
Nevertheless, despite the fact that the number of parameters significantly increases, we find that the impact of marginalizing over the correlated noise parameters is modest. For the three-detector case, the degradation of the constraints on each SGWB parameter is by a factor of $\sim2.7$ at most. 
Adding one more detector (i.e., KAGRA), the constraints can be improved, and the degradation is as small as a factor of $\sim2.0$. The main reason of this small impact comes from the fact that there is no significant degeneracy between the SGWB parameters and newly introduced parameters for correlated magnetic noise.

\subsubsection{Simultaneous constraints on SGWB and magnetic noise parameters}
\label{subsec:simultaneous_parameter_constraints}

To clarify the last point mentioned in Sec.~\ref{subsec:parameter_estimation_w_correlated_noise},  Fig.~\ref{fig:Contour_Meyer_design} shows the expected error contours on both the SGWB parameters and the coupling parameters of the correlated noise (i.e., $\kappa_I$ and $\beta_I$), with the orientation angle $\psi_I$ marginalized over. In the three-detector case depicted as blue shaded colors,  we see a rather large statistical error on the coupling parameters of correlated magnetic noise, and strong degeneracies among the coupling parameters of the different detectors are observed (e.g., $\kappa_{\rm H}$ and $\kappa_{\rm V}$, $\beta_{\rm H}$ and $\beta_{\rm V}$). Nevertheless, looking at the error contours for the SGWB parameter and the coupling parameter, shown in the bottom and second from the bottom panels, no significant degeneracy is found, and a degradation of the constraining power on the SGWB parameters remains insignificant compared to a large uncertainty in the coupling parameters. A part of the reasons comes from the structure of the Fisher matrix, in which all of the off-diagonal components associated with both the correlated noise and SGWB parameters, e.g., $F_{\kappa_I,\Omega_{\rm gw,0}}$, $F_{\beta_I,\Omega_{\rm gw,0}}$, and $F_{\psi_I,\Omega_{\rm gw,0}}$, become vanishing. Importantly, adding one more detector (KAGRA), the available number of cross-correlation statistics increases and improves substantially the constraining power on the parameters of correlated magnetic noise. As a result, the statistical errors of the coupling parameters shown in shaded orange in Fig.~\ref{fig:Contour_Meyer_design}, are dramatically reduced, and the parameter degeneracies between coupling parameters are broken, leading to a further improvement in the constraint on the SGWB.  

Here we comment on a similar analysis made in Ref.~\cite{Meyers:2020qrb}, in which the impact of correlated magnetic noise is estimated based on the Bayesian statistical inference. Our setup of the Fisher matrix analysis is close to their setup, but has several differences in detail. Apart from the methodology, one important difference is the orientation angle to characterize the directional dependence of the coupling with magnetic fields. Another major difference comes from the specification of the detector setup. Because of these, the results of forecast presented here cannot be directly compared to those obtained in Ref.~\cite{Meyers:2020qrb}. However, if one reexamines the parameter estimation errors using the Fisher matrix analysis with almost the same setup as in Ref.~\cite{Meyers:2020qrb}, they are turned out to be quantitatively consistent with their errors. Hence it is verified that our Fisher matrix analysis is relevant and reliable despite several assumptions and the model we adopted. This point is  discussed in detail in Appendix \ref{appendix:comparison_Myers_etal}.

\begin{figure*}[htb!]
\begin{center}
\includegraphics[width=18cm]{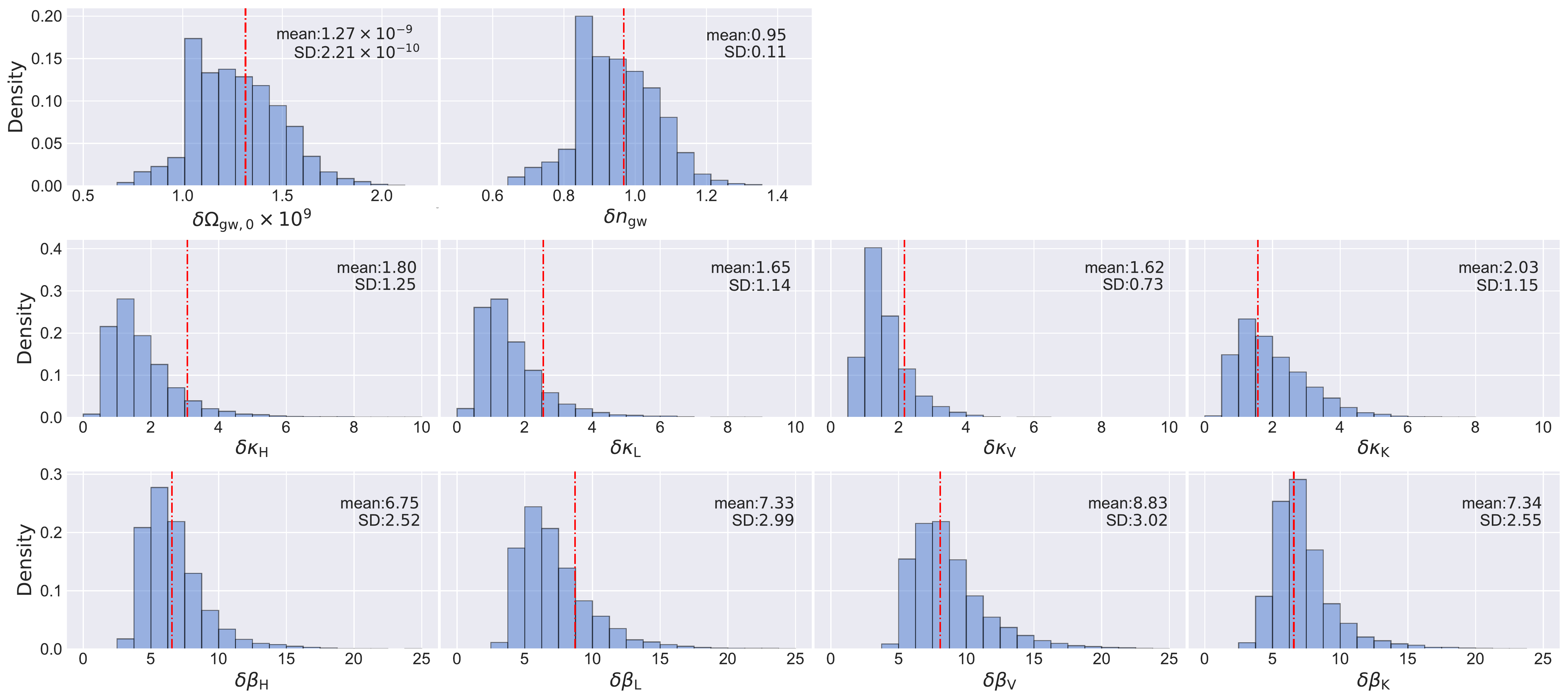}\\
\caption{Probability distribution of one-dimensional marginalized error for each parameter. Varying the fiducial value of orientation angles $\psi_I$ randomly, we repeat the Fisher matrix analysis in the four-detector case, and derive the one-dimensional marginalized error on each parameter. The resultant error distribution obtained from $10,000$ realizations is shown as the histogram, with the mean and the standard deviation (SD) of their distributions indicated in each panel. For reference, the vertical red dot-dashed lines represent the estimated size of one-dimensional errors obtained from the fiducial setup of $\psi_I$ in Table \ref{tab:magnetic-parameters}. }
\label{fig:projection-angles}
\end{center}
\end{figure*}

\subsubsection{Impact of coupling strengths}

So far we have evaluated the parameter estimation errors for the detector networks with the fiducial values of the magnetic coupling parameters. However, the magnetic coupling parameters would change easily if the status of instruments is changed. In Appendix~\ref{app:coupling-parameters}, we investigate the impact of the variation of coupling strength on the parameter estimation errors. We summarize the results below.

First we scale a set of the magnetic coupling strengths by a factor from the fiducial values. In this case, the parameter estimation errors of $\Omega_{{\rm gw},0}$ and $ n_{\rm gw}$ are insensitive to the scaling of the magnetic noise. It is because the correlation between the SGWB parameters and the magnetic coupling parameters are rather weak as mentioned in the previous section. 

Next we fix the coupling strengths for some detectors to the fiducial values and vary those for other detectors. As shown in Appendix~\ref{app:coupling-parameters}, again the measurement errors of $\Omega_{{\rm gw},0}$ and $ n_{\rm gw}$ are almost insensitive to the magnetic coupling strength. However, only in the case with the fiducial values of $\kappa_{\rm H}$ and $\kappa_{\rm L}$ and the larger values of $\kappa_{\rm V}=\kappa_{\rm K}$, the measurement errors of the SGWB parameters are slightly affected. This is because the SGWB parameters are determined predominantly from the LIGO pair, HL, while the magnetic coupling strengths are determined from the detector pair of Virgo and KAGRA separately. That is, the independent detector pairs play different roles in determining the parameters. 

From the results above, we conclude that the variation of the magnetic coupling strength hardly affects the parameter estimation error of $\Omega_{{\rm gw},0}$ and $ n_{\rm gw}$.

\subsubsection{Impact of orientation angles}
\label{subsubsec:orientation_angle}

Next we investigate the variation of the orientation angles $\psi_I$ kept fixed to those summarized in Table \ref{tab:magnetic-parameters}. As it has been shown in Refs.~\cite{2017PhRvD..96b2004H,2019PhRvD.100h2001H}, the parameter $\psi_I$ can change not only the amplitude but also the spectral shape of the magnetic noise spectrum. If we set a different value of $\psi_I$, forecast results may differ from those presented in Secs.~\ref{subsec:parameter_estimation_w_correlated_noise} and  \ref{subsec:simultaneous_parameter_constraints}. 

To see how the setup of the orientation angles affects the forecast results on the SGWB and magnetic coupling parameters, we randomly generate a set of the orientation angles. Then for each set of the orientation angles, the Fisher matrix analysis is performed in the four-detector case. With $10^4$ random realizations, we obtain the ensemble of one-dimensional marginalized errors of each parameter, and the results are plotted in  Fig.~\ref{fig:projection-angles} as the histograms of error distributions. Note that the parameters except for the orientation angles are kept fixed to fiducial values. Thus, the scatters of the errors directly reflect the sensitivity to the variation of the orientation angles. Looking at the coupling parameters of the correlated magnetic noise, their error distributions tend to have long tails toward a larger value of $\delta\kappa_I$ and $\delta\beta_I$. By contrast, the error distributions of the SGWB parameters are nearly symmetric, and have a broad peak around the mean values. Apart from an apparent impression on the trends in each parameter, the standard deviations of these error distributions is always smaller than their mean values, as indicated in each panel of Fig.~\ref{fig:projection-angles}. This means that
a quantitative impact of the orientation angles on the size of the errors is not significant. In particular, for the SGWB parameters, the variations of derived constraints remain at a level of $20\sim40$\%. Furthermore, compared with the results in Fig.~\ref{fig:Contour_Meyer_design},  we see that our fiducial setup of the orientation angles gives the mean or median values of the error distributions for each parameter, depicted as red dot-dashed vertical lines. This indicates that forecast results in   Secs.~\ref{subsec:parameter_estimation_w_correlated_noise} and  \ref{subsec:simultaneous_parameter_constraints} are not atypical but rather considered as a vanilla example.

\subsection{Systematic biases ignoring correlated magnetic noise}
\label{subsec:systematic_biases}

\begin{figure}[t]
\begin{center}
\includegraphics[width=8.5cm]{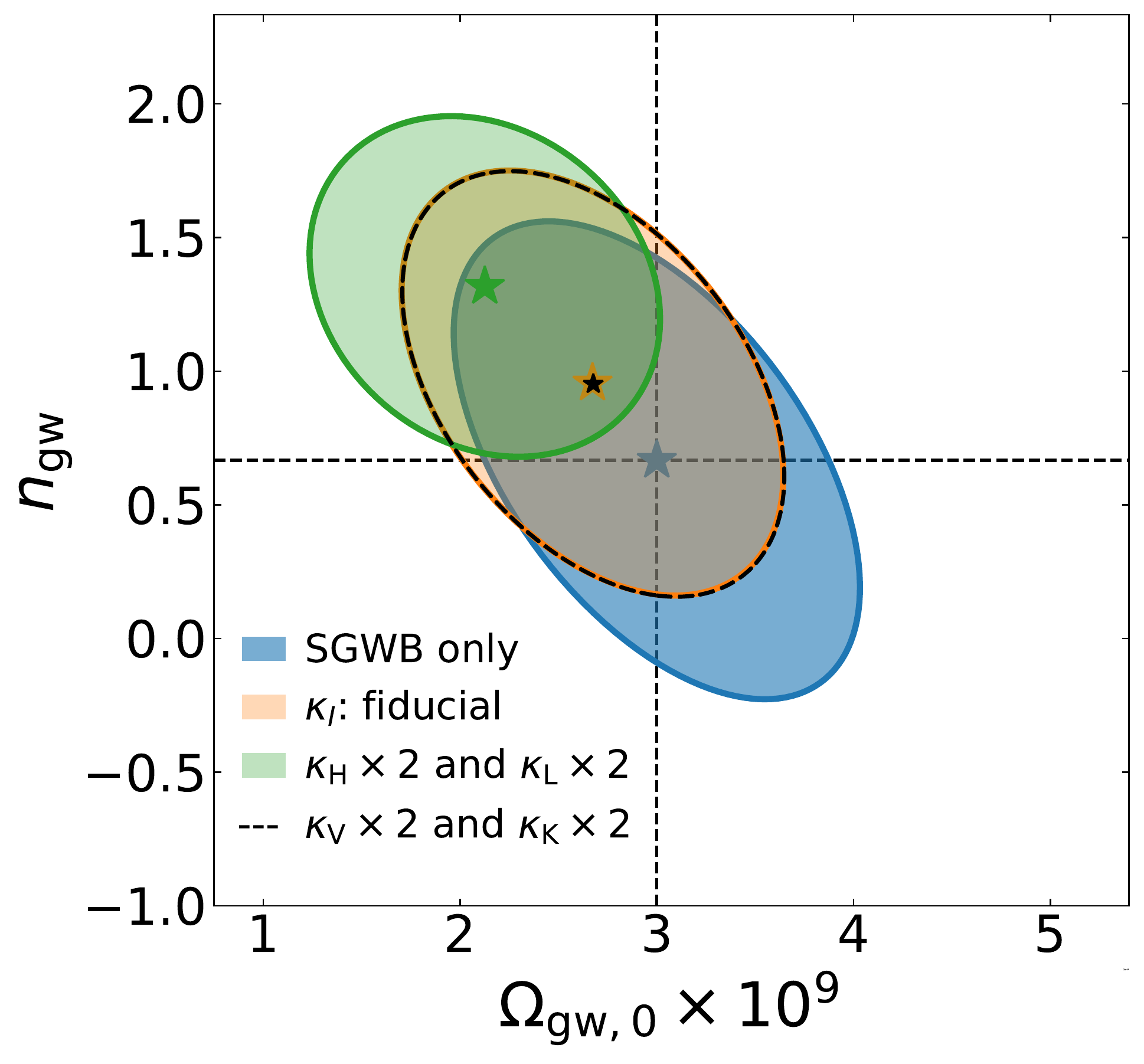}
\caption{Systematic biases for the best-fit values and uncertainty of $\Omega_{\rm gw, 0}$ and $n_{\rm gw}$. Ignoring the correlated magnetic noise, the biases for the SGWB parameters are estimated, and the errors around their biased best-fit values, as shown in star symbols, are computed in the case of the four-detector network (HLVK), based on the formalism in Sec.~\ref{subsec:Fisher-bias}.  While the orange shaded contour represents the result adopting the fiducial setup, the green shaded contour indicates the case when we increase the coupling parameter $\kappa_I$ for LIGO HL detectors by a factor of 2. On the other hand, the dashed ellipse shows the case if we instead increase $\kappa_I$ for Virgo and KAGRA by a factor of 2, which yields the result indistinguishable from the one for the fiducial setup. For reference, the blue shaded contour indicates the $1\sigma$ statistical error in the absence of correlated magnetic noise, which is identical to the blue shaded contour in Fig.~\ref{fig:Error_Contour_wo_Mag}.
}
\label{fig:bias}
\end{center}
\end{figure}

Forecast results presented so far assume that we have a good model to characterize the correlated magnetic noise. In this subsection, we examine the (in some sense extreme) case that we could not quantitatively model the correlated noise spectral features, and estimate the systematic impact of ignoring them on the best-fit values of the SGWB parameters. Based on the formalism described in Sec.~\ref{subsec:Fisher-bias}, we consider the four-detector case, and evaluate the systematic biases on the estimated parameters of a SGWB, i.e., $\Delta\Omega_{\rm gw,0}$ and $\Delta n_{\rm gw}$. Also, the statistical errors around the biased parameters are computed, and the results are compared with those in the absence of correlated magnetic noise, shown in Fig.~\ref{fig:bias}. 

Figure~\ref{fig:bias} shows that the SGWB parameters tend to be, overall, biased toward large and small values of $n_{\rm gw}$ and $\Omega_{\rm gw,0}$, respectively. The main reason for these biases comes from the fact that the correlation signal from a SGWB described as $U_{IJ}^{\rm gw}$ for the LIGO HL pair, 
which is the best sensitive pair of detectors, has an opposite sign to $U_{IJ}^{\rm mag}$ (see Fig.~\ref{fig:spectral-densities}). Thus, the sum of the correlation terms, $U_{IJ}^{\rm gw}+U_{IJ}^{\rm mag}$, is reduced due to a partial cancellation. As we saw from Fig.~\ref{fig:spectral-densities}, this cancellation becomes significant at lower frequencies, and thus the cross-correlation statistics, $U_{IJ}^{\rm gw}+U_{IJ}^{\rm mag}$, has an apparently different profile from the true spectrum of the SGWB.
As a result, we can erroneously detect a SGWB signal having a bluer tilted spectrum with a smaller amplitude than expected.

Nevertheless, we find that the systematic bias is not significant for our current setup with the fiducial couplings, depicted as the orange shaded contour, and is well within 1$\sigma$ statistical errors in the absence of correlated magnetic noise (blue contour). An important point is that the biases are rather sensitive to the choice of coupling parameter $\kappa_I$  for the LIGO detectors. For illustrative purposes, we increase the parameters $\kappa_{\rm H}$ and $\kappa_{\rm L}$ by a factor of 2, and estimate the size of systematic biases. Then, the systematic bias ignoring the correlated noise becomes serious, as shown in shaded green in Fig.~\ref{fig:bias}, and the best-fit values of $\Omega_{\rm gw,0}$ and $n_{\rm gw}$, as indicated by the green star symbol, are now almost at the boundary of the 1$\sigma$ error contour. On the other hand, if we instead increase the coupling parameters for Virgo and KAGRA ($\kappa_{\rm V}$ and $\kappa_{\rm K}$) by a factor of two, the systematic bias remains almost unchanged, and as shown in the black dashed contour, the result coincides with the one in the fiducial case. This is mainly because the LIGO HL pair has the best sensitivity to the SGWB, indicating that an unbiased parameter estimation ignoring the correlated noise requires a sufficiently suppressed coupling especially for the most sensitive detector pair. Nevertheless, a network detection including both Virgo and KAGRA is still powerful in simultaneously measuring both the SGWB and correlated noise, and is immune to the impact of correlated magnetic noise, as discussed in Sec.~\ref{subsec:parameter_estimation_w_correlated_noise}. In this respect, the effort to mitigate the couplings to the magnetic fields is indispensable for each detector toward a robust detection of SGWBs.

\section{Conclusion} 
\label{sec:conclusion}

In this paper, we have studied the impact of the correlated noise, arising especially from the global magnetic fields present in the Earth-ionosphere cavity, on the detection of SGWBs with a network of ground-based detectors. Such global magnetic fields are known as the Schumann resonances. Their couplings with mirror systems in laser interferometers can be a significant correlated noise source at a low-frequency band of $f\lesssim 50$\,Hz. The induced correlated noise between spatially separated detectors can make the detection of SGWBs difficult with the cross-correlation method. In order to detect a SGWB in the presence of correlated noise, one simple way is to model the correlated noise and to estimate the model parameters of the SGWB and correlated magnetic noise simultaneously. 

Here, we systematically investigated this issue based on the Fisher matrix analysis. Adopting an analytical model developed by Refs.~\cite{2017PhRvD..96b2004H,2019PhRvD.100h2001H} as a theoretical template of correlated magnetic noise, we estimated how much the constraints on the SGWB parameters, i.e., the amplitude ($\Omega_{\rm gw,0}$) and spectral index ($n_{\rm gw}$) of the energy density parameter, are degraded in the presence of the correlated magnetic noise for the various setups of detector networks and the model parameters of correlated magnetic noise. 

As explicitly demonstrated in Appendix \ref{appendix:comparison_Myers_etal}, our Fisher matrix calculations can give a result quantitatively consistent with Ref.~\cite{Meyers:2020qrb}, which demonstrates a joint parameter estimation in the framework of Bayesian statistical inference.
A crucial difference from Ref.~\cite{Meyers:2020qrb} in the present analysis is that we took into account 1 more degree of freedom to characterize the correlated magnetic noise at each detector site for a joint parameter estimation. It describes the directional coupling between a magnetic field and a detector (mirror displacement). We call this parameter the orientation angle and denote it by $\psi_I$. Accordingly, together with the SGWB parameters and the correlated noise parameters, the number of free parameters is in total $11$ for three detectors, and $14$ for four detectors. As it has been demonstrated in detail in Refs.~\cite{2017PhRvD..96b2004H,2019PhRvD.100h2001H}, the orientation angle $\psi_I$ can significantly change the frequency dependence of the magnetic coherence function $\gamma_{IJ}^{\rm mag}$ characterizing a detector response to the Schumann resonances. Since $\psi_I$ might not be determined precisely by monitoring the Schumann resonances with magnetometers, we allowed it to be free and paid special attention to the impact of this directional coupling on the joint parameter estimation of the SGWB and correlated noise. Furthermore, we have considered the parameter estimation of the SGWB parameters only, and quantified the systematic impact of ignoring the correlated magnetic noise on the best-fit values of the SGWB parameters. 

The important findings of our Fisher matrix analysis are summarized as follows:

\begin{itemize}
    \item Overall, there is no significant degeneracy between the parameters of a SGWB and correlated noise. Hence despite the fact that the number of parameters significantly increases, the impact of marginalizing over the correlated noise parameters is modest. This is consistent with Ref.~\cite{Meyers:2020qrb} and is indeed the cases with the additional parameter $\psi_I$ to marginalize over. With the design sensitivities of the second-generation detectors shown in Fig.~\ref{fig:noise-curves} and the correlated noise parameters summarized in Table~\ref{tab:magnetic-parameters}, the degradation of the constraints on the SGWB parameters is by a factor of $\sim2.7$ at most for the three-detector case (LIGO HL and Virgo).  Adding KAGRA, the constraints are improved, and the degradation is as small as a factor of $\sim2.0$.
    \item The forecast results mentioned above are robust against the variation of correlated noise parameters. When the magnetic coupling strengths are varied from the fiducial values up to a factor of 2.5, the parameter estimation errors of $\Omega_{{\rm gw},0}$ and $ n_{\rm gw}$ are hardly affected ($\lesssim 3\%$)  (see Appendix~\ref{app:coupling-parameters}). This is simply because the correlation between the SGWB parameters and the magnetic coupling parameters are rather weak. On the other hand, the variation of orientation angle $\psi_I$ is found to give a non-negligible change in the forecast results, but such an impact remains small as far as the SGWB parameters are concerned, and is at a level of $20\sim40$\% (Fig.~\ref{fig:projection-angles}).  
    \item Ignoring the correlated noise parameters in parameter estimation, the SGWB parameters are biased in general. However, for detectors having the design sensitivity and the correlated noise with the fiducial coupling, the impact of the systematic bias is found to be insignificant. Note, however, that this conclusion is rather sensitive to the coupling parameters of the LIGO detectors. Increasing their coupling strengths by a factor of 2 for each detector would lead to a serious systematic bias, and the best-fit values are largely shifted to the boundary of $1\sigma$ error contour (green shaded region of Fig.~\ref{fig:bias}). 
\end{itemize}

Throughout the paper, we have focused on the second-generation detectors and estimated the impacts of their correlated magnetic noise based on the Fisher matrix analysis. However, the formalism considered  here is fairly general, and can be applied to the third-generation detectors such as Cosmic Explorer \cite{Evans:2016mbw} and Einstein Telescope \cite{ET:2020}, for which the correlated magnetic noise would be potentially one of the most significant noise sources in a low frequency band. 
Our results suggest that if we can properly model the correlated noise in a parametric manner and marginalize its model parameters through the joint parameter estimation, the impact of correlated magnetic noise becomes less significant even in the case of the third-generation detectors. In this respect, a crucial point would be to accurately model the correlated magnetic noise. In this paper, we have used the analytical model developed in Refs.~\cite{2017PhRvD..96b2004H, 2019PhRvD.100h2001H}. Despite several assumptions and simplification, the model quantitatively explains the measured trends of the global coherence of the Schumann resonances \cite{2014PhRvD..90b3013T}. Nevertheless, toward a more accurate description of correlated noise, a concurrent data monitoring of magnetic fields with magnetometers would be rather crucial and helpful not only to improve the sensitivity to SGWBs but also to mitigate the nonstationary effects of the Schumann resonances. With a multiple set of the magnetic field data monitored in different directions, one can elucidate the directional couplings of detectors to the Schumann resonances. An experimental study with actual datasets would be an important next step, and we will discuss it elsewhere.

\begin{acknowledgments}
This work was supported in part by MEXT/JSPS KAKENHI Grants 
No. JP21K03580 (Y.H.), Grants No. JP20H05861 and No. JP21H01081 (A.T.), and Grants No. JP19H01894 and No. JP20H04726 (A.N.). A.T. acknowledges the support from JST AIP Acceleration Research Grant No. JP20317829, Japan. A.N. was also supported by research grants from Inamori Foundation.
\end{acknowledgments}

\appendix

\section{Magnetic field correlation, $M_{IJ}$}
\label{appendix:analytic_formula}


In this Appendix, we present the analytic expressions for the function $M_{IJ}$ given in Eq.~(\ref{eq:S_B_Fourier}). 

As discussed in Refs.~\cite{2017PhRvD..96b2004H, 2019PhRvD.100h2001H}, 
the function $M_{IJ}$ describes a detector response to the global magnetic fields, and characterizes both the strength and coherence of the magnetic fields at two separated detectors. To derive an analytical model of it, Ref.~\cite{2017PhRvD..96b2004H} has invoked assumptions and simplifications below, keeping essential properties of the Schumann resonances: 

\begin{enumerate}
\item Schumann resonances are described by a  superposition of the axisymmetric transverse magnetic (TM) modes of the Earth-ionosphere cavity generated by continuous and stationary random lightning excitation sources, whose distribution is isotropic.
The random amplitude of each TM mode, ${\widetilde B}$, is statistically characterized by the power spectrum $P_{\rm B}(f)$ in the following power-law form:
\begin{align}
 P_{\rm B}(f)=A \left(\frac{f}{10 {\rm Hz}}\right)^{-\alpha}\,,
\label{eq:mg_spectrum}
\end{align}
where we set in this paper the amplitude $A$ and slope $\alpha$ to $0.665$\,pT$^{2}$\,/Hz and $\alpha = 0.943$, respectively, with which we show in Fig.~\ref{fig:magnetic-spectrum} that the function $M$ resembles Fig.~2 of Ref.~\cite{Meyers:2020qrb}.\footnote{To be precise, these parameters are determined so as to get closer to Fig.~2 of Ref.~\cite{Meyers:2020qrb} when summing up the Schumann resonance modes up to $\ell=5$, convolving also with the line shape functions $E_\ell$. }

\item  The frequency-dependent coherence of the global magnetic field between two detectors is represented by a sum of discrete Schumann resonance modes primarily peaked at the frequencies $f_\ell\equiv c/(2\pi\,R_\oplus) \sqrt{\ell(\ell+1)}$, where $R_\oplus$, $c$ and $\ell$ are the radius of the Earth, light velocity and a positive integer, respectively. However, due to an imperfect conductivity at the boundary of the Earth-ionosphere cavity system, each of their modes is convolved with a line shape function $E_\ell(f)$ having a shifted peak around $f'_\ell$, given by
\begin{align}
|E_\ell(f)|^{2} \propto \frac{1}{(f-f'_{\ell})^{2}+\{f_{\ell}/(2{\cal Q})\}^{2}}\,,
\label{eq:spectral_form}
\end{align}
where the frequency $f'_\ell$ is determined empirically to match the observed resonance frequencies, and is related to $f_\ell$ through $f'_\ell=0.78f_\ell$ \cite{1998clel.book.....J}, which yields $8.3$, $14.3$, and $20.2$Hz for the lowest three modes. The quantity ${\cal Q}$ is the quality factor, which we adopt ${\cal Q}=3.21$.\footnote{In our previous study \cite{2017PhRvD..96b2004H, 2019PhRvD.100h2001H},  ${\cal Q}$ is set to $5$, but with the value of ${\cal Q}=3.21$, we find that the global magnetic field spectrum $M$ reproduces reasonably well the behavior seen in Fig.~2 of Ref.~\cite{Meyers:2020qrb}.}
\end{enumerate}

Along the line of these assumptions, we can rewrite the function $M_{IJ}$ with
\begin{align}
 M_{I J}(f) =P_{\rm B}(f)\,\,\sum_{\ell} \frac{|E_\ell(f)|^{2}}{|E_\ell(f'_{\ell})|^{2}}\,
\gamma_{\ell}^{\rm B}(\widehat{\bm r}_{1},\widehat{\bm r}_{2})\,.
\label{eq:mij_2}
\end{align}
In the above, an important building block is the function $\gamma_{\ell}^{\rm B}$ given as a function of unit vectors $\widehat{\bm r}_{I}$ and $\widehat{\bm r}_{J}$, which respectively point from the Earth's center to the {\it I}th and {\it J}th detector positions. This function specifically characterizes the coherence of global magnetic fields for each Schumann resonance mode, which sensitively depends on the geometrical configuration of a detector pair. The analytical expression of this function is given by \cite{2017PhRvD..96b2004H}
\begin{align}
&\gamma_{\ell}^{\rm B}(\widehat{\bm r}_{I},\widehat{\bm r}_{J}) =
  \frac{2\ell+1}{2\pi}\frac{(\ell-1)!}{(\ell+1)!} \nonumber\\
& \times \Bigl[
F_\ell(\mu)\,\left\{\mu\,(\widehat{\bm X}_I \cdot \widehat{\bm X}_J)
-(\widehat{\bm r}_{J} \cdot \widehat{\bm X}_I) \, (\widehat{\bm r}_{I} \cdot \widehat{\bm X}_J) \right\}
\nonumber \\
&
-G_\ell(\mu)\,\left\{(\widehat{\bm r}_I \times \widehat{\bm r}_J) \cdot \widehat{\bm X}_I \right\}
\, \left\{(\widehat{\bm r}_I \times \widehat{\bm r}_J) \cdot \widehat{\bm X}_J \right\}
\Bigr],
\label{eq:Gamma}
\end{align}
with the functions $F_\ell$ and $G_\ell$ defined below:
\begin{align}
&F_\ell(\mu) = -\frac{4\pi}{2\ell+1}\left[(\ell+1)\mathcal{P}_{\ell+1}(\mu)-\frac{\mu}{\sqrt{1-\mu^{2}}}\mathcal{P}_{\ell+1}^{1}(\mu)\right],
\label{eq:fl}
\\
&G_\ell(\mu) =  -\frac{4\pi}{2\ell+1}\,
\Biggl[(\ell+1)(\ell+2)\, \frac{\mu}{1-\mu^2}\, \mathcal{P}_{\ell+1}(\mu)
\nonumber \\
& \quad\quad\quad
+\frac{\ell-(\ell+2) \mu^{2}}{(1-\mu^{2})^{3/2}}\,\mathcal{P}_{\ell+1}^{1}(\mu)\Biggr]\,,
\label{eq:gl}
\end{align}
Here the function $\mathcal{P}_{\ell}^{1}$ is the associated Legendre polynomials, and 
$\mu$ is the directional cosine between the unit vectors $\widehat{\bm r}_I$ and $\widehat{\bm r}_J$, i.e., $\mu=\cos(\widehat{\bm r}_I\cdot\widehat{\bm r}_J)$. Notice that the function $\gamma_\ell^{\rm B}$ involves other unit vectors $\widehat{\bm X}_I$ and $\widehat{\bm X}_J$, which describe the directional dependence of the coupling with magnetic fields and lie at the tangent plane on the Earth at {\it I}th and {\it J}th detector positions, respectively [see Eq.~(\ref{eq:conv_noise})].

Given the analytical expression for the function $M_{IJ}$ above, 
in the main text, we decompose it into a product of two functions given by [see Eq.~(\ref{eq:mij})]
\begin{align}
& M_{IJ}(f) =
 M(f)\,\displaystyle{ \gamma_{IJ}^{\rm mag}(f) }.
\nonumber
\end{align}
Comparing the expression at Eq.~(\ref{eq:mij_2}) with the above form, the magnetic field spectrum $M(f)$ and the coherence function $\gamma_{IJ}^{\rm mag}(f)$ of our analytical model are given below: 
\begin{align}
& M(f)= P_{\rm B}(f)\,\,\displaystyle{\sum_{\ell} \frac{|E_\ell(f)|^{2}}{|E_\ell(f'_{\ell})|^{2}}}, 
\label{eq:power-spectrum}\\
&\displaystyle{ \gamma_{IJ}^{\rm mag}(f) =
\frac{ \sum_{\ell}\frac{|E_\ell(f)|^{2}}{|E_\ell(f'_{\ell})|^{2}}\,
\gamma_{\ell}^{\rm B}(\widehat{\bm r}_{I},\widehat{\bm r}_{J})}
{\sum_{\ell}\frac{|E_\ell(f)|^{2}}{|E_\ell(f'_{\ell})|^{2}}}}\,.
\label{eq:coherence}
\end{align}
These analytic expressions are used in the Fisher matrix analysis in Sec.~\ref{sec:results}, and Appendixes \ref{appendix:comparison_Myers_etal} and \ref{app:coupling-parameters}.

\begin{figure}[t]
\begin{center}
\includegraphics[width=8.5cm]{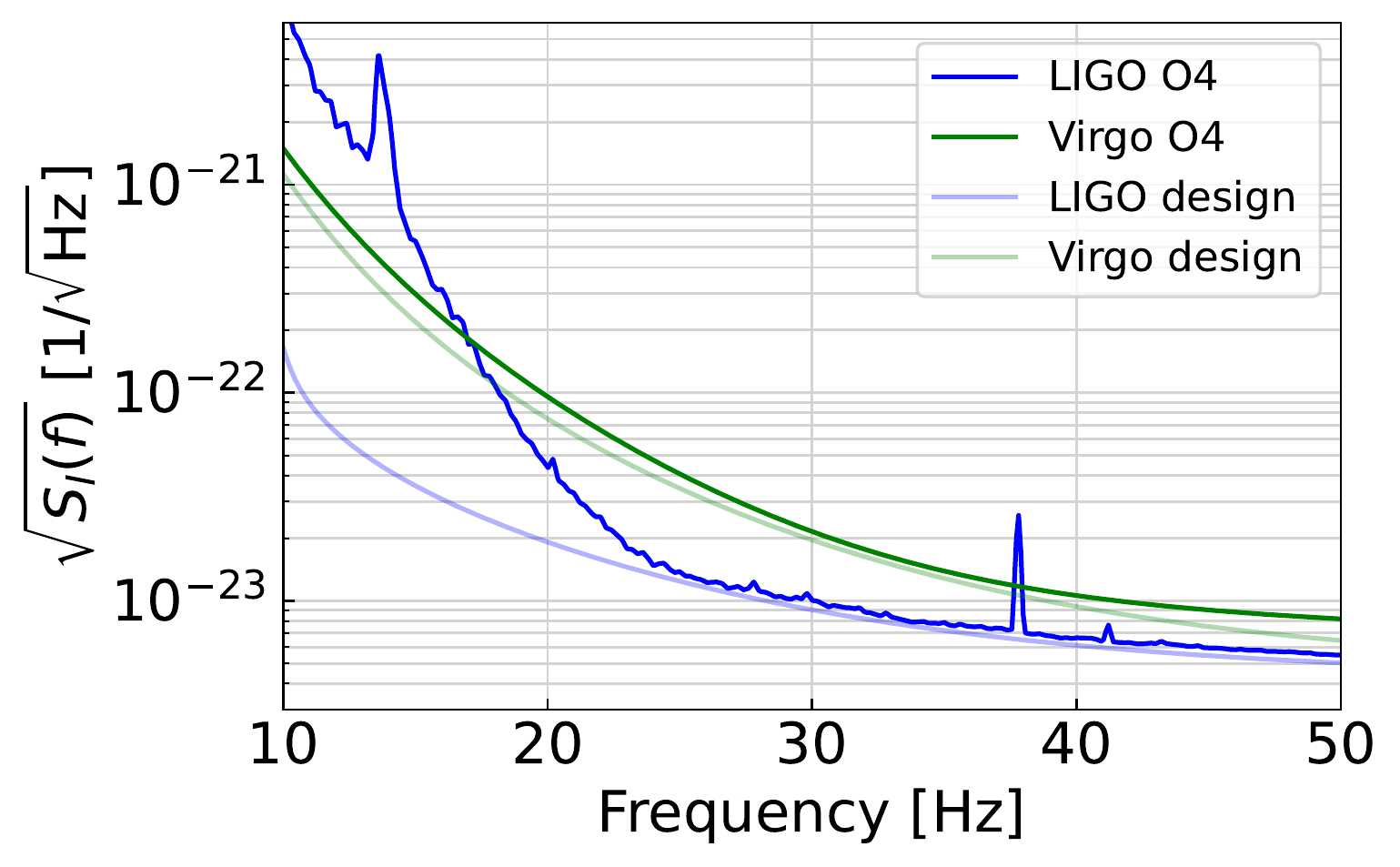}
\caption{Square root of the instrumental noise power spectral density, $\sqrt{S_{I}(f)}$, for LIGO (blue) and Virgo (green) adopting the sensitivity of O4.
For comparison, light blue and green curves represent the cases for the design sensitivity, identical to the blue and green lines in Fig.~\ref{fig:noise-curves}.}
\label{fig:noise-curves_update}
\end{center}
\end{figure}

\begin{figure*}[htb!]
\begin{center}
\includegraphics[width=17cm]{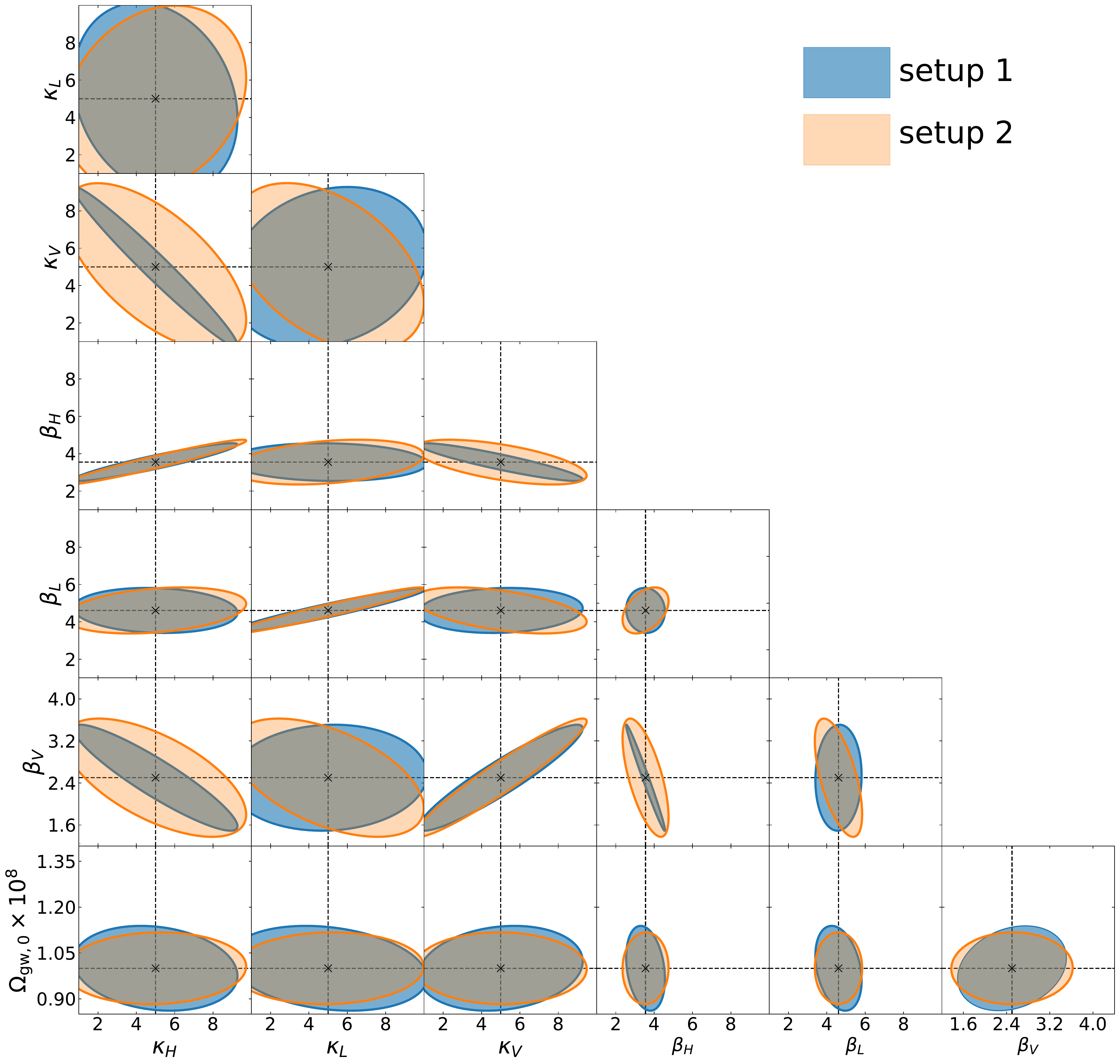}\\
\caption{
Expected error contours on both the SGWB and correlated noise parameters from joint parameter estimation, similar to what has been done in Ref.~\cite{Meyers:2020qrb}. Here, we consider the network of three detectors, HLV, and adopt the strong coupling parameters for $\kappa_I$ and $\beta_I$ listed in Table~\ref{tab:magnetic-parameters_setup2}. Fixing the spectral index $n_{\rm gw}$ to $2/3$ and excluding it and orientation angles from the Fisher matrix, we performed the Fisher matrix calculations with the amplitude of SGWB $\Omega_{\rm gw,0}=10^{-8}$. Forecast results are then plotted as the two-dimensional $1\sigma$ error on each pair of SGWB and coupling parameters. The blue and orange shaded contours respectively represent the results adopting the orientation angles of setups 1 and 2 as fiducial values (see Table~\ref{tab:magnetic-parameters_setup2}). 
}
\label{fig:Contour_Meyers}
\end{center}
\end{figure*}

\section{Comparison to Meyers {\it et al.}~\cite{Meyers:2020qrb}}
\label{appendix:comparison_Myers_etal}

In the main text of this paper, we considered a network of detectors, each of which has the design sensitivity, including KAGRA, depicted as solid curves in Fig.~\ref{fig:noise-curves}. 
In this Appendix, we compare our Fisher matrix calculations with those obtained by Ref~.\cite{Meyers:2020qrb}, in which the authors performed the Bayesian statistical inference with the three detectors (i.e., LIGO HL and Virgo), adopting the O4 sensitivity curves~\cite{LVK:2022}, as shown in  Fig.~\ref{fig:noise-curves_update}.

\begin{figure}[tb!]
\begin{center}
\includegraphics[width=8.5cm]{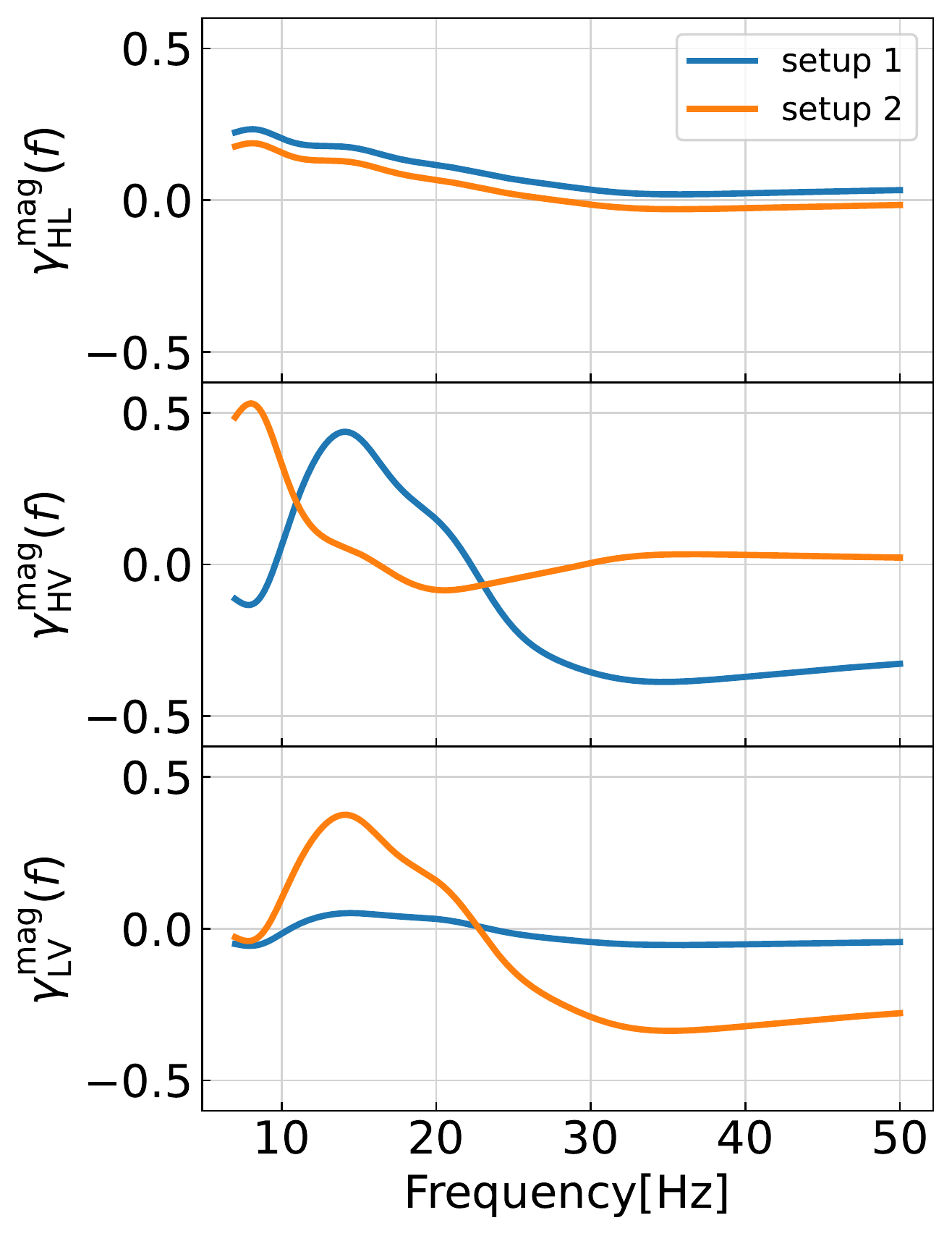}
\caption{Frequency dependence of magnetic coherence functions used in the Fisher matrix analysis in Appendix \ref{appendix:comparison_Myers_etal}. From top to bottom, results for LIGO HL pair, LIGO Hanford and Virgo, and LIGO Livingston and Virgo are shown. In each panel, blue and orange curves are the magnetic coherence functions adopting the orientation angles of the setup 1 and 2, respectively
(see Table~\ref{tab:magnetic-parameters_setup2}).  
}
\label{fig:magnetic-coherence_setup2}
\end{center}
\end{figure}

To compute the Fisher matrix with the setup similar to Ref.~\cite{Meyers:2020qrb}, we consider the strong coupling parameters, $(\kappa_I,\,\beta_I)$, for the coupling function with the magnetic fields, summarized in Table \ref{tab:magnetic-parameters_setup2}. Also, the fiducial value of the SGWB amplitude is changed to $\Omega_{\rm gw,0}=10^{-8}$, with the spectral index of $n_{\rm gw}=2/3$ kept fixed. Further, for consistency with the analysis in Ref.~\cite{Meyers:2020qrb}, the orientation angles of the correlated magnetic noise, $\psi_I$, are all fixed and excluded from the Fisher matrix components.
As a result, the number of free parameters to be determined is reduced to $7$ in the three-detector case (i.e., $\theta_a=\{\Omega_{\rm gw,0},\, \kappa_{\rm H/L/V}, \,\beta_{\rm H/L/V}\}$). Finally, in order to get closer to the analysis in  Ref.~\cite{Meyers:2020qrb}, we impose the Gaussian prior to the coupling parameters with the dispersion of $\sigma_\kappa=5$ and $\sigma_\beta=5$. Strictly, the prior information imposed in Ref.~\cite{Meyers:2020qrb} is not Gaussian but uniformly distributed in the range, and they also considered a log-uniform prior distribution for the SGWB amplitude. Nevertheless, as we see below, these differences would not drastically change the conclusion at least at a qualitative level.

Figure~\ref{fig:Contour_Meyers} presents the results of our Fisher matrix calculations. Here we plot the two-dimensional $1\sigma$ error contours for each pair of parameters, as similarly shown in Fig.~5 of Ref.~\cite{Meyers:2020qrb} (see particularly their blue contours). In each panel, setting the orientation angles to those adopted in the main text (setup 1), which we list in Table~\ref{tab:magnetic-parameters} (see also Table~\ref{tab:magnetic-parameters_setup2}), forecast results are shown in blue. Note that with this setup, the magnetic coherence functions remain the same as shown in Fig.~\ref{fig:magnetic-coherence} (see also Fig.~\ref{fig:magnetic-coherence_setup2}). 
To see how our forecast results are sensitive to the fiducial setup of the correlated magnetic noise, we also consider in Fig.~\ref{fig:Contour_Meyers} another setup of the coherence functions, adopting a different set of orientation angles listed in Table \ref{tab:magnetic-parameters_setup2}  (setup 2).
The latter setup is 
chosen by comparing the coherence functions $\gamma_{IJ}^{\rm mag}$ of our analytical model with those measured from magnetometers in Ref.~\cite{Meyers:2020qrb} in the frequency range of $20-40\,$Hz, where the detectors with the O4 sensitivity curves become most sensitive to the SGWB. The results are shown in orange contours. 
In Fig.~\ref{fig:magnetic-coherence_setup2}, the magnetic coherence functions with the orientation angle of setup 2 are shown in orange curves.

Comparing our Fisher forecast results with the $1\sigma$ errors obtained from the Bayesian analysis in Ref.\cite{Meyers:2020qrb} 
(dark blue shaded contours in their Fig.~5), 
we see that the size of statistical errors are rather consistent with each other. Although the resultant constraints on $\kappa_I$ are mostly determined by the prior information we impose, we still see that the slopes of the couplings, $\beta_I$, are well constrained with statistical errors smaller than the prior distributions. Further, the SGWB amplitude $\Omega_{\rm gw,0}$ is determined unambiguously, with the one-dimensional error down to $\delta\Omega_{\rm gw,0}\sim0.1\times10^{-8}$ and no notable difference between the results from the Fisher matrix and the Bayesian statistical analysis in Ref.~\cite{Meyers:2020qrb}.
More importantly, parameter degeneracies as indicated by the orientations of error eclipses, in particular for the results in the setup 2, almost coincide with what were obtained in Ref.~\cite{Meyers:2020qrb}, apart from their parameter distributions shifted from the fiducial values. 
Hence, we conclude that despite several assumptions for simplifying a model, our Fisher matrix analysis can reliably reproduce the results quantitatively consistent with those of the full Bayesian statistical analysis.

\begin{table}[tb]
\caption{Magnetic coupling parameters and orientation angles (in units of radians) used in the Fisher matrix analysis in Appendix \ref{appendix:comparison_Myers_etal}.}
\label{tab:magnetic-parameters_setup2}
\begin{center}
\begin{tabular}{|l|c|c|c|c|}
\hline

\hline
Detectors & $\kappa_I$ & $\beta_I$ & $\psi_I$ (setup 1) & $\psi_I$ (setup 2)\\
\hline\hline
LIGO (Hanford) & 5.0 & 3.55 & 5.97 & 1.04 \\
LIGO (Livingston) & 5.0 &  4.61 & 0.64 & 5.68 \\
Virgo & 5.0 & 2.50 & 1.12 & 6.00\\
\hline
\end{tabular}
\end{center}
\end{table}

\section{Variation of the magnetic coupling strengths}
\label{app:coupling-parameters}

In the main text, we chose the magnetic coupling parameters listed in Table~\ref{tab:magnetic-parameters} as our fiducial values. However, in the future observation, these values will be certainly changed. In this Appendix, we analytically investigate how the different values of the coupling parameters affect the results of parameter estimation in some limiting cases.

To derive the parameter estimation errors analytically, we assume that the noise power spectral density, $S_I$, is identical with each other in the sensitive frequency band of detectors. For notational convenience, we denote the derivatives in the Fisher matrix in Eq.~(\ref{eq:Fisher-matrix}) as
\begin{align}
G_{IJ} &\equiv \frac{\partial U_{IJ} }{\partial \Omega_{{\rm gw},0}} = \frac{U_{IJ}^{\rm gw}}{\Omega_{{\rm gw},0}} \;, \\ B_{IJ} &\equiv \frac{1}{\kappa_J}\frac{\partial U_{IJ} }{\partial \kappa_I} = M_{IJ} \left( \frac{f}{10\,{\rm Hz}} \right)^{-\beta_I-\beta_J} \;, \\
Q_{IJ}^{GG} &\equiv 2 T_{\rm obs}\int_0^{\infty} \frac{G_{IJ}^2}{S_I^2} df \;, \\ 
Q_{IJ}^{GB} &\equiv 2 T_{\rm obs} \int_0^{\infty} \frac{G_{IJ} B_{IJ}}{S_I^2} df \;, \\ 
Q_{IJ}^{BB} &\equiv 2 T_{\rm obs} \int_0^{\infty} \frac{B_{IJ}^2}{S_I^2} df \;, 
\end{align}
Note that $G_{IJ}$ and $B_{IJ}$ are symmetric about the exchange of the indices.

\subsection{Case 1: Three detectors}
\label{app:case1}

We consider three detectors, HLV, and take $\Omega_{{\rm gw},0}$, $\kappa_{\rm H}$, $\kappa_{\rm L}$, $\kappa_{\rm V}$ as the free parameters in the Fisher matrix. Then the Fisher matrix for the HL detector pair is   
\begin{equation}
\mathbf{F}_{\rm HL} = 
\left(
\begin{array}{cccc} 
Q_{\rm HL}^{\rm GG} & \kappa_{\rm L} Q_{\rm HL}^{\rm GB} & \kappa_{\rm H} Q_{\rm HL}^{\rm GB} & 0 \vspace{.3em} \\
& \kappa_{\rm L}^2 Q_{\rm HL}^{\rm BB} & \kappa_{\rm H}\kappa_{\rm L} Q_{\rm HL}^{\rm BB} & 0 \vspace{.3em} \\
& & \kappa_{\rm L}^2 Q_{\rm HL}^{\rm BB} & 0 \vspace{.3em} \\
& & & 0 \vspace{.3em}
\end{array} 
\right) \;.
\end{equation}
The total Fisher matrix is given by $\mathbf{F}=\mathbf{F}_{\rm HL} + \mathbf{F}_{\rm HV} + \mathbf{F}_{\rm LV}$. The parameter estimation error of $\Omega_{{\rm gw},0}$ is computed as 
\begin{align}
\delta \Omega_{{\rm gw},0} &= F_{\Omega_{{\rm gw},0}\Omega_{{\rm gw},0}}^{-1/2} \nonumber \\
&= \left[ \sum_{(I,J)} Q_{\rm IJ}^{GG} - \frac{(Q_{\rm IJ}^{GB})^2}{Q_{\rm IJ}^{BB}} \right]^{-1/2} \;,
\end{align}
where the sum is taken for all detector pairs. Interestingly, the $\Omega_{{\rm gw},0}$ error is independent of the magnetic coupling strength, $\kappa_I$. It is verified numerically that the independence of $\kappa_I$ holds even when $n_{\rm gw}$ is included as an additional free parameter as shown in Fig.~\ref{fig:kappa_dependence1}. As the sensitivity to a SGWB is predominantly determined by the HL pair, the $\Omega_{{\rm gw},0}$ error is plotted as a function of $\kappa_{\rm V}$.

\subsection{Case 2: Four detectors with $\kappa_{\rm H}=\kappa_{\rm L}=\kappa_{\rm V}=\kappa_1$ and $\kappa_{\rm K}=\kappa_2$}
\label{app:case2}

As the sensitivity to a SGWB is predominantly determined by the HL pair and the magnetic coupling of KAGRA is still uncertain, we consider the case of $\kappa_{\rm H}=\kappa_{\rm L}=\kappa_{\rm V}=\kappa_1$ and $\kappa_{\rm K}=\kappa_2$. We assume that the frequency integrals are classified into two classes: 
\begin{align}
Q_{\rm HL}^{GB} &= Q_{\rm HV}^{GB} = Q_{\rm LV}^{GB} \;, \\
Q_{\rm HK}^{GB} &= Q_{\rm LK}^{GB} = Q_{\rm VK}^{GB} \;, 
\end{align}
and similarly for $Q_{\rm IJ}^{BB}$. In addition, we neglect GW correlation signals from the detector pairs other than HL, which has a dominant contribution. The parameter estimation error of $\Omega_{{\rm gw},0}$ is 
\begin{align}
\delta \Omega_{{\rm gw},0} &= F_{\Omega_{{\rm gw},0}\Omega_{{\rm gw},0}}^{-1/2} \nonumber \\
&= \left[ Q_{\rm HL}^{GG} - 3 \frac{(Q_{\rm HL}^{GB})^2}{Q_{\rm HL}^{BB}} - 3 \frac{(Q_{\rm HK}^{GB})^2}{Q_{\rm HK}^{BB}} \right]^{-1/2} \;.
\end{align}
Again the $\Omega_{{\rm gw},0}$ error is independent of the magnetic coupling strength, $\kappa_I$. In the left panel of Fig.~\ref{fig:kappa_dependence1}, the result is verified numerically for the case with fixed $\beta_I$ and $\psi_I$.

\subsection{Case 3: Four detectors with $\kappa_{\rm H}=\kappa_{\rm L}=\kappa_1$ and $\kappa_{\rm V}=\kappa_{\rm K}=\kappa_2$}
\label{app:case3}

\begin{figure*}[t!]
\begin{center}
\includegraphics[width=8.5cm]{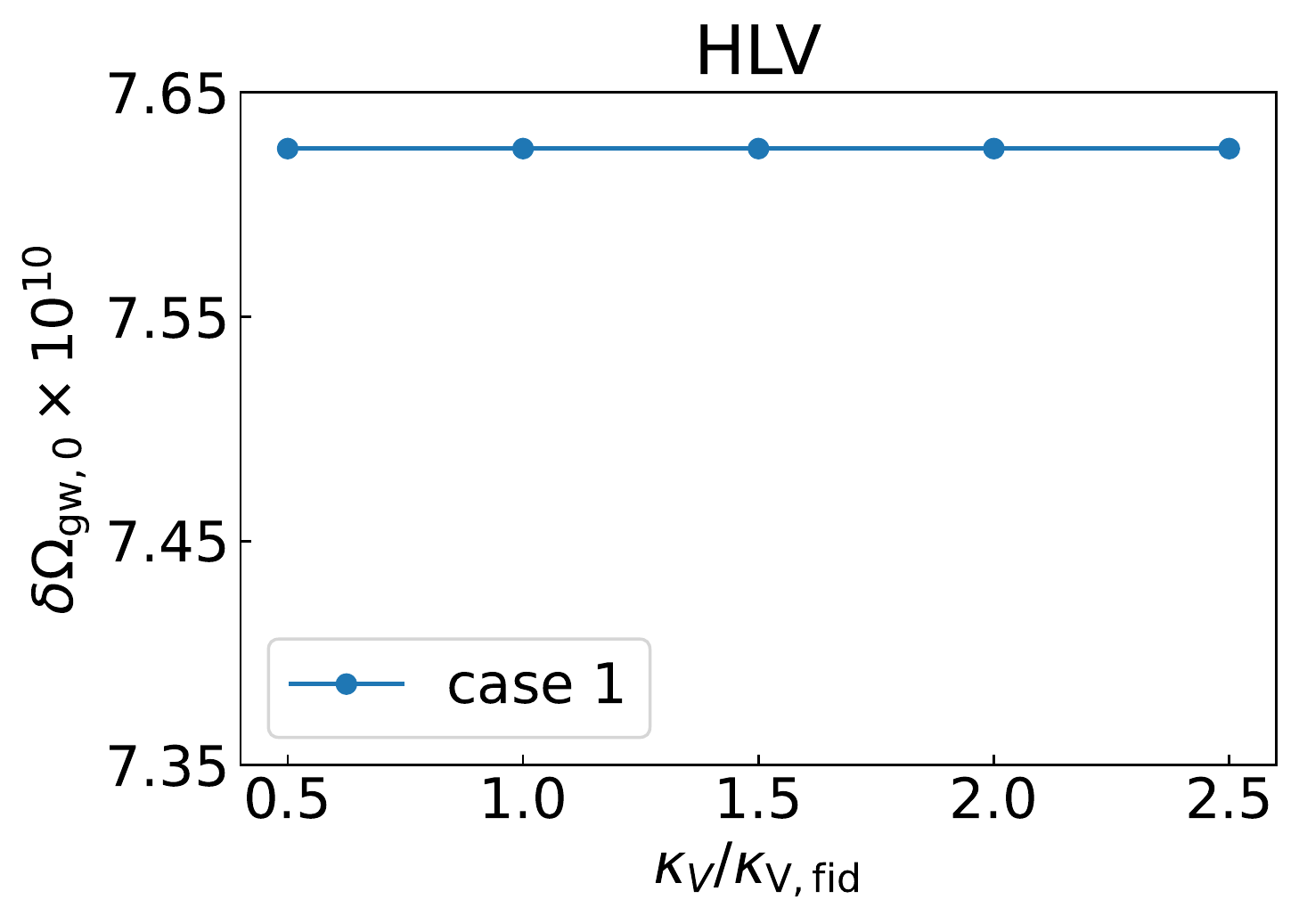}
\includegraphics[width=8.5cm]{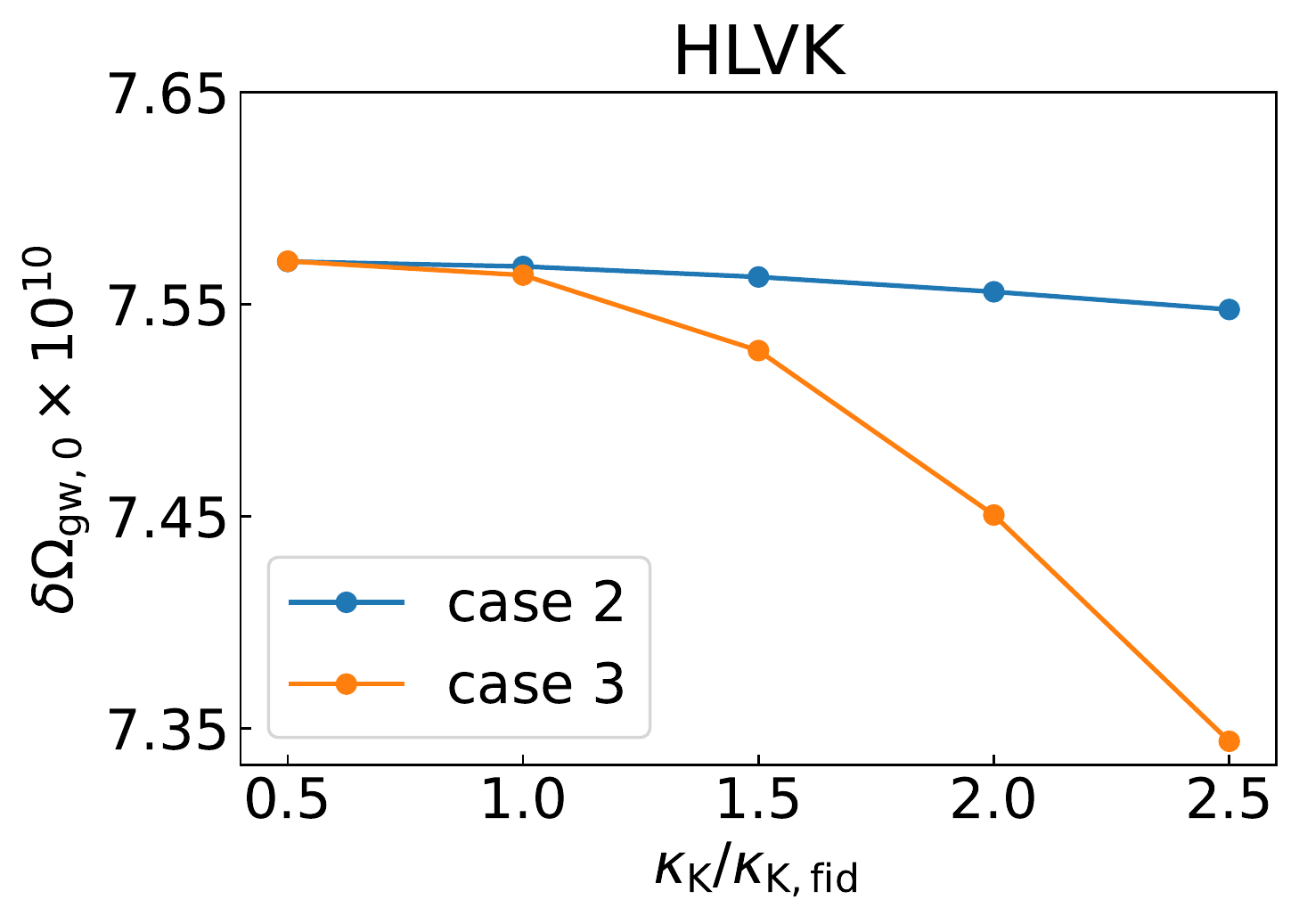}
\caption{Impact of the coupling strength for the three-detector (HLV) and four-detector (HLVK) cases, shown respectively in left and right panels. The horizontal axes in the left and right panels respectively represent the coupling strength of Virgo and KAGRA, normalized by their fiducial values. Then the estimated size of the marginalized error on the SGWB amplitude, $\delta\Omega_{\rm gw,0}$, is plotted for case 1 in the left panel, and cases 2 and 3 in the right panels (see Appendixes \ref{app:case1}, \ref{app:case2} and \ref{app:case3}). In all cases, the parameter $\beta_I$ and orientation angle $\psi_I$ are fixed to their fiducial values in Table~\ref{tab:magnetic-parameters}, and are excluded from the Fisher matrix.
}
\label{fig:kappa_dependence1}
\end{center}
\end{figure*}

Similar to case 2, we may also consider the situation such that two of four detectors are the most sensitive to the SGWB, and others are not. We set the coupling strengths to  $\kappa_{\rm H}=\kappa_{\rm L}=\kappa_1$ and $\kappa_{\rm V}=\kappa_{\rm K}=\kappa_2$. We assume further that some of the frequency integrals give the same contribution: 
\begin{equation}
Q_{\rm HV}^{GB} = Q_{\rm LV}^{GB} = Q_{\rm HK}^{GB} = Q_{\rm LK}^{GB} \;. \end{equation}
and similarly for $Q_{\rm IJ}^{BB}$. In addition, we neglect GW correlation signals from the detector pairs other than HL. The parameter estimation error of $\Omega_{{\rm gw},0}$ is 
\begin{widetext}
\begin{align}
\delta \Omega_{{\rm gw},0} &= F_{\Omega_{{\rm gw},0}\Omega_{{\rm gw},0}}^{-1/2} \nonumber \\
&= \left[ Q_{\rm HL}^{GG} - \bar{Q}^{BB} \left\{ \frac{(Q_{\rm HL}^{GB})^2}{Q_{\rm HL}^{BB}} \left(\frac{1}{Q_{\rm VK}^{BB}} + \frac{\kappa_r^2}{Q_{\rm HV}^{BB}} \right) + \frac{(Q_{\rm VK}^{GB})^2}{Q_{\rm VK}^{BB}} \left(\frac{\kappa_r^2}{Q_{\rm HV}^{BB}} + \frac{\kappa_r^4}{Q_{\rm HL}^{BB}} \right) + 4 \frac{(Q_{\rm HV}^{GB})^2}{Q_{\rm HV}^{BB}} \left(\frac{1}{Q_{\rm VK}^{BB}} + \frac{\kappa_r^4}{Q_{\rm HL}^{BB}} \right) \right. \right. \nonumber \\ 
&\qquad \left. \left. + 4 \kappa_r \left( \frac{Q_{\rm HV}^{GB}Q_{\rm VK}^{GB}}{Q_{\rm HV}^{BB}Q_{\rm VK}^{BB}} - \frac{\kappa_r}{2} \frac{Q_{\rm HL}^{GB}Q_{\rm VK}^{GB}}{Q_{\rm HL}^{BB}Q_{\rm VK}^{BB}} + \kappa_r^2 \frac{Q_{\rm HV}^{GB}Q_{\rm HL}^{GB}}{Q_{\rm HV}^{BB}Q_{\rm HL}^{BB}} \right)  \right\} \right]^{-1/2} \;,
\label{eq:OmegaGW-err-case3}
\end{align}
\end{widetext}
\begin{equation}
\bar{Q}^{BB} \equiv \left( \frac{1}{Q_{\rm VK}^{BB}} + \frac{\kappa_r^2}{Q_{\rm HV}^{BB}} + \frac{\kappa_r^4}{Q_{\rm HL}^{BB}}  \right)^{-1} \;.
\end{equation}
The $\Omega_{{\rm gw},0}$ error depends on the ratio of the magnetic coupling strength, $\kappa_r$. It is verified numerically in the right panel of Fig.~\ref{fig:kappa_dependence1} that the $\Omega_{{\rm gw},0}$ error slightly decreases as $\kappa_r$ increases when $\beta_I$ and $\psi_I$ are fixed.

If $\kappa_r$ is set to $1$ (all the coupling strengths are the same) and all $Q_{\rm IJ}^{GB}$ are equal to $Q_{\rm HL}^{GB}$, Eq.~(\ref{eq:OmegaGW-err-case3}) is reduced to
\begin{equation}
\delta \Omega_{{\rm gw},0} = \left[ Q_{\rm HL}^{GG} - 6 \frac{(Q_{\rm HL}^{GB})^2}{Q_{\rm HL}^{BB}} \right]^{-1/2} \;.
\end{equation}
The dependence of the $\Omega_{{\rm gw},0}$ error on the magnetic coupling strength disappears.


\begin{thebibliography}{38}%
\makeatletter
\providecommand \@ifxundefined [1]{%
 \@ifx{#1\undefined}
}%
\providecommand \@ifnum [1]{%
 \ifnum #1\expandafter \@firstoftwo
 \else \expandafter \@secondoftwo
 \fi
}%
\providecommand \@ifx [1]{%
 \ifx #1\expandafter \@firstoftwo
 \else \expandafter \@secondoftwo
 \fi
}%
\providecommand \natexlab [1]{#1}%
\providecommand \enquote  [1]{``#1''}%
\providecommand \bibnamefont  [1]{#1}%
\providecommand \bibfnamefont [1]{#1}%
\providecommand \citenamefont [1]{#1}%
\providecommand \href@noop [0]{\@secondoftwo}%
\providecommand \href [0]{\begingroup \@sanitize@url \@href}%
\providecommand \@href[1]{\@@startlink{#1}\@@href}%
\providecommand \@@href[1]{\endgroup#1\@@endlink}%
\providecommand \@sanitize@url [0]{\catcode `\\12\catcode `\$12\catcode
  `\&12\catcode `\#12\catcode `\^12\catcode `\_12\catcode `\%12\relax}%
\providecommand \@@startlink[1]{}%
\providecommand \@@endlink[0]{}%
\providecommand \url  [0]{\begingroup\@sanitize@url \@url }%
\providecommand \@url [1]{\endgroup\@href {#1}{\urlprefix }}%
\providecommand \urlprefix  [0]{URL }%
\providecommand \Eprint [0]{\href }%
\providecommand \doibase [0]{http://dx.doi.org/}%
\providecommand \selectlanguage [0]{\@gobble}%
\providecommand \bibinfo  [0]{\@secondoftwo}%
\providecommand \bibfield  [0]{\@secondoftwo}%
\providecommand \translation [1]{[#1]}%
\providecommand \BibitemOpen [0]{}%
\providecommand \bibitemStop [0]{}%
\providecommand \bibitemNoStop [0]{.\EOS\space}%
\providecommand \EOS [0]{\spacefactor3000\relax}%
\providecommand \BibitemShut  [1]{\csname bibitem#1\endcsname}%
\let\auto@bib@innerbib\@empty
\bibitem [{\citenamefont {Aasi}\ \emph {et~al.}(2015)\citenamefont {Aasi},
  \citenamefont {Abbott}, \citenamefont {Abbott}, \citenamefont {Abbott},
  \citenamefont {Abernathy}, \citenamefont {Ackley}, \citenamefont {Adams},
  \citenamefont {Adams}, \citenamefont {Addesso} \emph {et~al.}}]{LIGO1}%
  \BibitemOpen
  \bibfield  {author} {\bibinfo {author} {\bibfnamefont {J.}~\bibnamefont
  {Aasi}}, \bibinfo {author} {\bibfnamefont {B.~P.}\ \bibnamefont {Abbott}},
  \bibinfo {author} {\bibfnamefont {R.}~\bibnamefont {Abbott}}, \bibinfo
  {author} {\bibfnamefont {T.}~\bibnamefont {Abbott}}, \bibinfo {author}
  {\bibfnamefont {M.~R.}\ \bibnamefont {Abernathy}}, \bibinfo {author}
  {\bibfnamefont {K.}~\bibnamefont {Ackley}}, \bibinfo {author} {\bibfnamefont
  {C.}~\bibnamefont {Adams}}, \bibinfo {author} {\bibfnamefont
  {T.}~\bibnamefont {Adams}}, \bibinfo {author} {\bibfnamefont
  {P.}~\bibnamefont {Addesso}},  \emph {et~al.},\ }\href {\doibase
  10.1088/0264-9381/32/7/074001} {\bibfield  {journal} {\bibinfo  {journal}
  {Classical and Quantum Gravity}\ }\textbf {\bibinfo {volume} {32}},\ \bibinfo
  {pages} {074001} (\bibinfo {year} {2015})}\BibitemShut {NoStop}%
\bibitem [{\citenamefont {Harry}(2010)}]{LIGO2}%
  \BibitemOpen
  \bibfield  {author} {\bibinfo {author} {\bibfnamefont {G.~M.}\ \bibnamefont
  {Harry}},\ }\href {\doibase 10.1088/0264-9381/27/8/084006} {\bibfield
  {journal} {\bibinfo  {journal} {Classical and Quantum Gravity}\ }\textbf
  {\bibinfo {volume} {27}},\ \bibinfo {pages} {084006} (\bibinfo {year}
  {2010})}\BibitemShut {NoStop}%
\bibitem [{\citenamefont {Acernese}\ \emph {et~al.}(2015)\citenamefont
  {Acernese}, \citenamefont {Agathos}, \citenamefont {Agatsuma}, \citenamefont
  {Aisa}, \citenamefont {Allemandou}, \citenamefont {Allocca}, \citenamefont
  {Amarni}, \citenamefont {Astone}, \citenamefont {Balestri}, \citenamefont
  {Ballardin} \emph {et~al.}}]{Virgo}%
  \BibitemOpen
  \bibfield  {author} {\bibinfo {author} {\bibfnamefont {F.}~\bibnamefont
  {Acernese}}, \bibinfo {author} {\bibfnamefont {M.}~\bibnamefont {Agathos}},
  \bibinfo {author} {\bibfnamefont {K.}~\bibnamefont {Agatsuma}}, \bibinfo
  {author} {\bibfnamefont {D.}~\bibnamefont {Aisa}}, \bibinfo {author}
  {\bibfnamefont {N.}~\bibnamefont {Allemandou}}, \bibinfo {author}
  {\bibfnamefont {A.}~\bibnamefont {Allocca}}, \bibinfo {author} {\bibfnamefont
  {J.}~\bibnamefont {Amarni}}, \bibinfo {author} {\bibfnamefont
  {P.}~\bibnamefont {Astone}}, \bibinfo {author} {\bibfnamefont
  {G.}~\bibnamefont {Balestri}}, \bibinfo {author} {\bibfnamefont
  {G.}~\bibnamefont {Ballardin}},  \emph {et~al.},\ }\href {\doibase
  10.1088/0264-9381/32/2/024001} {\bibfield  {journal} {\bibinfo  {journal}
  {Classical and Quantum Gravity}\ }\textbf {\bibinfo {volume} {32}},\ \bibinfo
  {pages} {024001} (\bibinfo {year} {2015})}\BibitemShut {NoStop}%
\bibitem [{\citenamefont {{Abbott}}\ \emph {et~al.}(2016)\citenamefont
  {{Abbott}} \emph {et~al.}}]{GW150914:detection}%
  \BibitemOpen
  \bibfield  {author} {\bibinfo {author} {\bibfnamefont {B.~P.}\ \bibnamefont
  {{Abbott}}} \emph {et~al.} (\bibinfo {collaboration} {LIGO Scientific
  Collaboration and Virgo Collaboration}),\ }\href {\doibase
  10.1103/PhysRevLett.116.061102} {\bibfield  {journal} {\bibinfo  {journal}
  {Physical Review Letters}\ }\textbf {\bibinfo {volume} {116}},\ \bibinfo
  {eid} {061102} (\bibinfo {year} {2016})},\ \Eprint
  {http://arxiv.org/abs/1602.03837} {arXiv:1602.03837 [gr-qc]} \BibitemShut
  {NoStop}%
\bibitem [{\citenamefont {Abbott}\ \emph {et~al.}(2019)\citenamefont {Abbott}
  \emph {et~al.}}]{LIGOScientific:2018mvr}%
  \BibitemOpen
  \bibfield  {author} {\bibinfo {author} {\bibfnamefont {B.}~\bibnamefont
  {Abbott}} \emph {et~al.} (\bibinfo {collaboration} {LIGO Scientific,
  Virgo}),\ }\href {\doibase 10.1103/PhysRevX.9.031040} {\bibfield  {journal}
  {\bibinfo  {journal} {Phys. Rev. X}\ }\textbf {\bibinfo {volume} {9}},\
  \bibinfo {pages} {031040} (\bibinfo {year} {2019})},\ \Eprint
  {http://arxiv.org/abs/1811.12907} {arXiv:1811.12907 [astro-ph.HE]}
  \BibitemShut {NoStop}%
\bibitem [{\citenamefont {Abbott}\ \emph
  {et~al.}(2021{\natexlab{a}})\citenamefont {Abbott} \emph
  {et~al.}}]{LIGOScientific:2020ibl}%
  \BibitemOpen
  \bibfield  {author} {\bibinfo {author} {\bibfnamefont {R.}~\bibnamefont
  {Abbott}} \emph {et~al.} (\bibinfo {collaboration} {LIGO Scientific,
  Virgo}),\ }\href {\doibase 10.1103/PhysRevX.11.021053} {\bibfield  {journal}
  {\bibinfo  {journal} {Phys. Rev. X}\ }\textbf {\bibinfo {volume} {11}},\
  \bibinfo {pages} {021053} (\bibinfo {year} {2021}{\natexlab{a}})},\ \Eprint
  {http://arxiv.org/abs/2010.14527} {arXiv:2010.14527 [gr-qc]} \BibitemShut
  {NoStop}%
\bibitem [{\citenamefont {Abbott}\ \emph
  {et~al.}(2021{\natexlab{b}})\citenamefont {Abbott} \emph
  {et~al.}}]{LIGOScientific:2021djp}%
  \BibitemOpen
  \bibfield  {author} {\bibinfo {author} {\bibfnamefont {R.}~\bibnamefont
  {Abbott}} \emph {et~al.} (\bibinfo {collaboration} {LIGO Scientific, VIRGO,
  KAGRA}),\ }\href@noop {} {\  (\bibinfo {year} {2021}{\natexlab{b}})},\
  \Eprint {http://arxiv.org/abs/2111.03606} {arXiv:2111.03606 [gr-qc]}
  \BibitemShut {NoStop}%
\bibitem [{\citenamefont {Abbott}\ \emph
  {et~al.}(2021{\natexlab{c}})\citenamefont {Abbott} \emph
  {et~al.}}]{KAGRA:2021kbb}%
  \BibitemOpen
  \bibfield  {author} {\bibinfo {author} {\bibfnamefont {R.}~\bibnamefont
  {Abbott}} \emph {et~al.} (\bibinfo {collaboration} {KAGRA, Virgo, LIGO
  Scientific}),\ }\href {\doibase 10.1103/PhysRevD.104.022004} {\bibfield
  {journal} {\bibinfo  {journal} {Phys. Rev. D}\ }\textbf {\bibinfo {volume}
  {104}},\ \bibinfo {pages} {022004} (\bibinfo {year} {2021}{\natexlab{c}})},\
  \Eprint {http://arxiv.org/abs/2101.12130} {arXiv:2101.12130 [gr-qc]}
  \BibitemShut {NoStop}%
\bibitem [{\citenamefont {{Christensen}}(1992)}]{1992PhRvD..46.5250C}%
  \BibitemOpen
  \bibfield  {author} {\bibinfo {author} {\bibfnamefont {N.}~\bibnamefont
  {{Christensen}}},\ }\href {\doibase 10.1103/PhysRevD.46.5250} {\bibfield
  {journal} {\bibinfo  {journal} {\prd}\ }\textbf {\bibinfo {volume} {46}},\
  \bibinfo {pages} {5250} (\bibinfo {year} {1992})}\BibitemShut {NoStop}%
\bibitem [{\citenamefont {{Allen}}\ and\ \citenamefont
  {{Romano}}(1999)}]{1999PhRvD..59j2001A}%
  \BibitemOpen
  \bibfield  {author} {\bibinfo {author} {\bibfnamefont {B.}~\bibnamefont
  {{Allen}}}\ and\ \bibinfo {author} {\bibfnamefont {J.~D.}\ \bibnamefont
  {{Romano}}},\ }\href {\doibase 10.1103/PhysRevD.59.102001} {\bibfield
  {journal} {\bibinfo  {journal} {\prd}\ }\textbf {\bibinfo {volume} {59}},\
  \bibinfo {eid} {102001} (\bibinfo {year} {1999})},\ \Eprint
  {http://arxiv.org/abs/gr-qc/9710117} {gr-qc/9710117} \BibitemShut {NoStop}%
\bibitem [{\citenamefont
  {{Schumann}}(1952{\natexlab{a}})}]{1952ZNatA...7..149S}%
  \BibitemOpen
  \bibfield  {author} {\bibinfo {author} {\bibfnamefont {W.~O.}\ \bibnamefont
  {{Schumann}}},\ }\href {\doibase 10.1515/zna-1952-0202} {\bibfield  {journal}
  {\bibinfo  {journal} {Zeitschrift Naturforschung Teil A}\ }\textbf {\bibinfo
  {volume} {7}},\ \bibinfo {pages} {149} (\bibinfo {year}
  {1952}{\natexlab{a}})}\BibitemShut {NoStop}%
\bibitem [{\citenamefont
  {{Schumann}}(1952{\natexlab{b}})}]{1952ZNatA...7..250S}%
  \BibitemOpen
  \bibfield  {author} {\bibinfo {author} {\bibfnamefont {W.~O.}\ \bibnamefont
  {{Schumann}}},\ }\href {\doibase 10.1515/zna-1952-3-404} {\bibfield
  {journal} {\bibinfo  {journal} {Zeitschrift Naturforschung Teil A}\ }\textbf
  {\bibinfo {volume} {7}},\ \bibinfo {pages} {250} (\bibinfo {year}
  {1952}{\natexlab{b}})}\BibitemShut {NoStop}%
\bibitem [{\citenamefont {{Thrane}}\ \emph {et~al.}(2013)\citenamefont
  {{Thrane}}, \citenamefont {{Christensen}},\ and\ \citenamefont
  {{Schofield}}}]{2013PhRvD..87l3009T}%
  \BibitemOpen
  \bibfield  {author} {\bibinfo {author} {\bibfnamefont {E.}~\bibnamefont
  {{Thrane}}}, \bibinfo {author} {\bibfnamefont {N.}~\bibnamefont
  {{Christensen}}}, \ and\ \bibinfo {author} {\bibfnamefont {R.~M.~S.}\
  \bibnamefont {{Schofield}}},\ }\href {\doibase 10.1103/PhysRevD.87.123009}
  {\bibfield  {journal} {\bibinfo  {journal} {\prd}\ }\textbf {\bibinfo
  {volume} {87}},\ \bibinfo {eid} {123009} (\bibinfo {year} {2013})},\ \Eprint
  {http://arxiv.org/abs/1303.2613} {arXiv:1303.2613 [astro-ph.IM]} \BibitemShut
  {NoStop}%
\bibitem [{\citenamefont {{Thrane}}\ \emph {et~al.}(2014)\citenamefont
  {{Thrane}}, \citenamefont {{Christensen}}, \citenamefont {{Schofield}},\ and\
  \citenamefont {{Effler}}}]{2014PhRvD..90b3013T}%
  \BibitemOpen
  \bibfield  {author} {\bibinfo {author} {\bibfnamefont {E.}~\bibnamefont
  {{Thrane}}}, \bibinfo {author} {\bibfnamefont {N.}~\bibnamefont
  {{Christensen}}}, \bibinfo {author} {\bibfnamefont {R.~M.~S.}\ \bibnamefont
  {{Schofield}}}, \ and\ \bibinfo {author} {\bibfnamefont {A.}~\bibnamefont
  {{Effler}}},\ }\href {\doibase 10.1103/PhysRevD.90.023013} {\bibfield
  {journal} {\bibinfo  {journal} {Physical Review D}\ }\textbf {\bibinfo
  {volume} {90}},\ \bibinfo {eid} {023013} (\bibinfo {year} {2014})},\ \Eprint
  {http://arxiv.org/abs/1406.2367} {arXiv:1406.2367 [astro-ph.IM]} \BibitemShut
  {NoStop}%
\bibitem [{\citenamefont {Coughlin}\ \emph {et~al.}(2016)\citenamefont
  {Coughlin} \emph {et~al.}}]{Coughlin:2016vor}%
  \BibitemOpen
  \bibfield  {author} {\bibinfo {author} {\bibfnamefont {M.~W.}\ \bibnamefont
  {Coughlin}} \emph {et~al.},\ }\href {\doibase 10.1088/0264-9381/33/22/224003}
  {\bibfield  {journal} {\bibinfo  {journal} {Class. Quant. Grav.}\ }\textbf
  {\bibinfo {volume} {33}},\ \bibinfo {pages} {224003} (\bibinfo {year}
  {2016})},\ \Eprint {http://arxiv.org/abs/1606.01011} {arXiv:1606.01011
  [gr-qc]} \BibitemShut {NoStop}%
\bibitem [{\citenamefont {Coughlin}\ \emph {et~al.}(2018)\citenamefont
  {Coughlin} \emph {et~al.}}]{Coughlin:2018tjc}%
  \BibitemOpen
  \bibfield  {author} {\bibinfo {author} {\bibfnamefont {M.~W.}\ \bibnamefont
  {Coughlin}} \emph {et~al.},\ }\href {\doibase 10.1103/PhysRevD.97.102007}
  {\bibfield  {journal} {\bibinfo  {journal} {Phys. Rev. D}\ }\textbf {\bibinfo
  {volume} {97}},\ \bibinfo {pages} {102007} (\bibinfo {year} {2018})},\
  \Eprint {http://arxiv.org/abs/1802.00885} {arXiv:1802.00885 [gr-qc]}
  \BibitemShut {NoStop}%
\bibitem [{\citenamefont {{Himemoto}}\ and\ \citenamefont
  {{Taruya}}(2017)}]{2017PhRvD..96b2004H}%
  \BibitemOpen
  \bibfield  {author} {\bibinfo {author} {\bibfnamefont {Y.}~\bibnamefont
  {{Himemoto}}}\ and\ \bibinfo {author} {\bibfnamefont {A.}~\bibnamefont
  {{Taruya}}},\ }\href {\doibase 10.1103/PhysRevD.96.022004} {\bibfield
  {journal} {\bibinfo  {journal} {\prd}\ }\textbf {\bibinfo {volume} {96}},\
  \bibinfo {eid} {022004} (\bibinfo {year} {2017})},\ \Eprint
  {http://arxiv.org/abs/1704.07084} {arXiv:1704.07084 [astro-ph.IM]}
  \BibitemShut {NoStop}%
\bibitem [{\citenamefont {{Himemoto}}\ and\ \citenamefont
  {{Taruya}}(2019)}]{2019PhRvD.100h2001H}%
  \BibitemOpen
  \bibfield  {author} {\bibinfo {author} {\bibfnamefont {Y.}~\bibnamefont
  {{Himemoto}}}\ and\ \bibinfo {author} {\bibfnamefont {A.}~\bibnamefont
  {{Taruya}}},\ }\href {\doibase 10.1103/PhysRevD.100.082001} {\bibfield
  {journal} {\bibinfo  {journal} {\prd}\ }\textbf {\bibinfo {volume} {100}},\
  \bibinfo {eid} {082001} (\bibinfo {year} {2019})},\ \Eprint
  {http://arxiv.org/abs/1908.10635} {arXiv:1908.10635 [astro-ph.IM]}
  \BibitemShut {NoStop}%
\bibitem [{\citenamefont {Kowalska-Leszczynska}\ \emph
  {et~al.}(2017)\citenamefont {Kowalska-Leszczynska} \emph
  {et~al.}}]{Kowalska-Leszczynska:2016low}%
  \BibitemOpen
  \bibfield  {author} {\bibinfo {author} {\bibfnamefont {I.}~\bibnamefont
  {Kowalska-Leszczynska}} \emph {et~al.},\ }\href {\doibase
  10.1088/1361-6382/aa60eb} {\bibfield  {journal} {\bibinfo  {journal} {Class.
  Quant. Grav.}\ }\textbf {\bibinfo {volume} {34}},\ \bibinfo {pages} {074002}
  (\bibinfo {year} {2017})},\ \Eprint {http://arxiv.org/abs/1612.01102}
  {arXiv:1612.01102 [astro-ph.IM]} \BibitemShut {NoStop}%
\bibitem [{\citenamefont {Washimi}\ \emph {et~al.}(2021)\citenamefont
  {Washimi}, \citenamefont {Yokozawa}, \citenamefont {Nakano}, \citenamefont
  {Tanaka}, \citenamefont {Kaihotsu}, \citenamefont {Mori},\ and\ \citenamefont
  {Narita}}]{Washimi:2021ogz}%
  \BibitemOpen
  \bibfield  {author} {\bibinfo {author} {\bibfnamefont {T.}~\bibnamefont
  {Washimi}}, \bibinfo {author} {\bibfnamefont {T.}~\bibnamefont {Yokozawa}},
  \bibinfo {author} {\bibfnamefont {M.}~\bibnamefont {Nakano}}, \bibinfo
  {author} {\bibfnamefont {T.}~\bibnamefont {Tanaka}}, \bibinfo {author}
  {\bibfnamefont {K.}~\bibnamefont {Kaihotsu}}, \bibinfo {author}
  {\bibfnamefont {Y.}~\bibnamefont {Mori}}, \ and\ \bibinfo {author}
  {\bibfnamefont {T.}~\bibnamefont {Narita}},\ }\href {\doibase
  10.1088/1748-0221/16/07/P07033} {\bibfield  {journal} {\bibinfo  {journal}
  {JINST}\ }\textbf {\bibinfo {volume} {16}},\ \bibinfo {pages} {P07033}
  (\bibinfo {year} {2021})},\ \Eprint {http://arxiv.org/abs/2103.06516}
  {arXiv:2103.06516 [gr-qc]} \BibitemShut {NoStop}%
\bibitem [{\citenamefont {Janssens}\ \emph {et~al.}(2023)\citenamefont
  {Janssens} \emph {et~al.}}]{Janssens:2022tdj}%
  \BibitemOpen
  \bibfield  {author} {\bibinfo {author} {\bibfnamefont {K.}~\bibnamefont
  {Janssens}} \emph {et~al.},\ }\href {\doibase 10.1103/PhysRevD.107.022004}
  {\bibfield  {journal} {\bibinfo  {journal} {Phys. Rev. D}\ }\textbf {\bibinfo
  {volume} {107}},\ \bibinfo {pages} {022004} (\bibinfo {year} {2023})},\
  \Eprint {http://arxiv.org/abs/2209.00284} {arXiv:2209.00284 [gr-qc]}
  \BibitemShut {NoStop}%
\bibitem [{\citenamefont {Akutsu}\ \emph {et~al.}(2021)\citenamefont {Akutsu}
  \emph {et~al.}}]{KAGRA_2021PTEP}%
  \BibitemOpen
  \bibfield  {author} {\bibinfo {author} {\bibfnamefont {T.}~\bibnamefont
  {Akutsu}} \emph {et~al.} (\bibinfo {collaboration} {KAGRA}),\ }\href
  {\doibase 10.1093/ptep/ptaa125} {\bibfield  {journal} {\bibinfo  {journal}
  {PTEP}\ }\textbf {\bibinfo {volume} {2021}},\ \bibinfo {pages} {05A101}
  (\bibinfo {year} {2021})},\ \Eprint {http://arxiv.org/abs/2005.05574}
  {arXiv:2005.05574 [physics.ins-det]} \BibitemShut {NoStop}%
\bibitem [{\citenamefont {Janssens}\ \emph {et~al.}(2021)\citenamefont
  {Janssens}, \citenamefont {Martinovic}, \citenamefont {Christensen},
  \citenamefont {Meyers},\ and\ \citenamefont
  {Sakellariadou}}]{Janssens:2021cta}%
  \BibitemOpen
  \bibfield  {author} {\bibinfo {author} {\bibfnamefont {K.}~\bibnamefont
  {Janssens}}, \bibinfo {author} {\bibfnamefont {K.}~\bibnamefont
  {Martinovic}}, \bibinfo {author} {\bibfnamefont {N.}~\bibnamefont
  {Christensen}}, \bibinfo {author} {\bibfnamefont {P.~M.}\ \bibnamefont
  {Meyers}}, \ and\ \bibinfo {author} {\bibfnamefont {M.}~\bibnamefont
  {Sakellariadou}},\ }\href {\doibase 10.1103/PhysRevD.104.122006} {\bibfield
  {journal} {\bibinfo  {journal} {Phys. Rev. D}\ }\textbf {\bibinfo {volume}
  {104}},\ \bibinfo {pages} {122006} (\bibinfo {year} {2021})},\ \bibinfo
  {note} {[Erratum: Phys.Rev.D 105, 109904 (2022)]},\ \Eprint
  {http://arxiv.org/abs/2110.14730} {arXiv:2110.14730 [gr-qc]} \BibitemShut
  {NoStop}%
\bibitem [{\citenamefont {Parida}\ \emph {et~al.}(2016)\citenamefont {Parida},
  \citenamefont {Mitra},\ and\ \citenamefont {Jhingan}}]{Parida:2015fma}%
  \BibitemOpen
  \bibfield  {author} {\bibinfo {author} {\bibfnamefont {A.}~\bibnamefont
  {Parida}}, \bibinfo {author} {\bibfnamefont {S.}~\bibnamefont {Mitra}}, \
  and\ \bibinfo {author} {\bibfnamefont {S.}~\bibnamefont {Jhingan}},\ }\href
  {\doibase 10.1088/1475-7516/2016/04/024} {\bibfield  {journal} {\bibinfo
  {journal} {JCAP}\ }\textbf {\bibinfo {volume} {04}},\ \bibinfo {pages} {024}
  (\bibinfo {year} {2016})},\ \Eprint {http://arxiv.org/abs/1510.07994}
  {arXiv:1510.07994 [astro-ph.CO]} \BibitemShut {NoStop}%
\bibitem [{\citenamefont {Martinovic}\ \emph {et~al.}(2021)\citenamefont
  {Martinovic}, \citenamefont {Meyers}, \citenamefont {Sakellariadou},\ and\
  \citenamefont {Christensen}}]{Martinovic:2020hru}%
  \BibitemOpen
  \bibfield  {author} {\bibinfo {author} {\bibfnamefont {K.}~\bibnamefont
  {Martinovic}}, \bibinfo {author} {\bibfnamefont {P.~M.}\ \bibnamefont
  {Meyers}}, \bibinfo {author} {\bibfnamefont {M.}~\bibnamefont
  {Sakellariadou}}, \ and\ \bibinfo {author} {\bibfnamefont {N.}~\bibnamefont
  {Christensen}},\ }\href {\doibase 10.1103/PhysRevD.103.043023} {\bibfield
  {journal} {\bibinfo  {journal} {Phys. Rev. D}\ }\textbf {\bibinfo {volume}
  {103}},\ \bibinfo {pages} {043023} (\bibinfo {year} {2021})},\ \Eprint
  {http://arxiv.org/abs/2011.05697} {arXiv:2011.05697 [gr-qc]} \BibitemShut
  {NoStop}%
\bibitem [{\citenamefont {Poletti}(2021)}]{Poletti:2021ytu}%
  \BibitemOpen
  \bibfield  {author} {\bibinfo {author} {\bibfnamefont {D.}~\bibnamefont
  {Poletti}},\ }\href {\doibase 10.1088/1475-7516/2021/05/052} {\bibfield
  {journal} {\bibinfo  {journal} {JCAP}\ }\textbf {\bibinfo {volume} {05}},\
  \bibinfo {pages} {052} (\bibinfo {year} {2021})},\ \Eprint
  {http://arxiv.org/abs/2101.02713} {arXiv:2101.02713 [gr-qc]} \BibitemShut
  {NoStop}%
\bibitem [{\citenamefont {Seto}\ and\ \citenamefont
  {Taruya}(2008)}]{Seto:2008sr}%
  \BibitemOpen
  \bibfield  {author} {\bibinfo {author} {\bibfnamefont {N.}~\bibnamefont
  {Seto}}\ and\ \bibinfo {author} {\bibfnamefont {A.}~\bibnamefont {Taruya}},\
  }\href {\doibase 10.1103/PhysRevD.77.103001} {\bibfield  {journal} {\bibinfo
  {journal} {Phys.Rev.}\ }\textbf {\bibinfo {volume} {D77}},\ \bibinfo {pages}
  {103001} (\bibinfo {year} {2008})},\ \Eprint {http://arxiv.org/abs/0801.4185}
  {arXiv:0801.4185 [astro-ph]} \BibitemShut {NoStop}%
\bibitem [{\citenamefont {Nishizawa}\ \emph {et~al.}(2009)\citenamefont
  {Nishizawa}, \citenamefont {Taruya}, \citenamefont {Hayama}, \citenamefont
  {Kawamura},\ and\ \citenamefont {Sakagami}}]{Nishizawa:2009bf}%
  \BibitemOpen
  \bibfield  {author} {\bibinfo {author} {\bibfnamefont {A.}~\bibnamefont
  {Nishizawa}}, \bibinfo {author} {\bibfnamefont {A.}~\bibnamefont {Taruya}},
  \bibinfo {author} {\bibfnamefont {K.}~\bibnamefont {Hayama}}, \bibinfo
  {author} {\bibfnamefont {S.}~\bibnamefont {Kawamura}}, \ and\ \bibinfo
  {author} {\bibfnamefont {M.-a.}\ \bibnamefont {Sakagami}},\ }\href {\doibase
  10.1103/PhysRevD.79.082002} {\bibfield  {journal} {\bibinfo  {journal}
  {Phys.Rev.}\ }\textbf {\bibinfo {volume} {D79}},\ \bibinfo {pages} {082002}
  (\bibinfo {year} {2009})},\ \Eprint {http://arxiv.org/abs/0903.0528}
  {arXiv:0903.0528 [astro-ph.CO]} \BibitemShut {NoStop}%
\bibitem [{\citenamefont {Meyers}\ \emph {et~al.}(2020)\citenamefont {Meyers},
  \citenamefont {Martinovic}, \citenamefont {Christensen},\ and\ \citenamefont
  {Sakellariadou}}]{Meyers:2020qrb}%
  \BibitemOpen
  \bibfield  {author} {\bibinfo {author} {\bibfnamefont {P.~M.}\ \bibnamefont
  {Meyers}}, \bibinfo {author} {\bibfnamefont {K.}~\bibnamefont {Martinovic}},
  \bibinfo {author} {\bibfnamefont {N.}~\bibnamefont {Christensen}}, \ and\
  \bibinfo {author} {\bibfnamefont {M.}~\bibnamefont {Sakellariadou}},\ }\href
  {\doibase 10.1103/PhysRevD.102.102005} {\bibfield  {journal} {\bibinfo
  {journal} {Phys. Rev. D}\ }\textbf {\bibinfo {volume} {102}},\ \bibinfo
  {pages} {102005} (\bibinfo {year} {2020})},\ \Eprint
  {http://arxiv.org/abs/2008.00789} {arXiv:2008.00789 [gr-qc]} \BibitemShut
  {NoStop}%
\bibitem [{\citenamefont {{Jackson}}(1998)}]{1998clel.book.....J}%
  \BibitemOpen
  \bibfield  {author} {\bibinfo {author} {\bibfnamefont {J.~D.}\ \bibnamefont
  {{Jackson}}},\ }\href@noop {} {\emph {\bibinfo {title} {Classical
  Electrodynamics, 3rd Edition}}}\ (\bibinfo  {publisher} {Wiley},\ \bibinfo
  {year} {1998})\ p.\ \bibinfo {pages} {374}\BibitemShut {NoStop}%
\bibitem [{\citenamefont {{Shoemaker}}(2010)}]{Shoemaker:2014}%
  \BibitemOpen
  \bibfield  {author} {\bibinfo {author} {\bibfnamefont {D.}~\bibnamefont
  {{Shoemaker}}},\ }\href {https://dcc.ligo.org/LIGO-T0900288-v3/public}
  {\bibfield  {journal} {\bibinfo  {journal}
  {https://dcc.ligo.org/LIGO-T0900288-v3/public}\ } (\bibinfo {year}
  {2010})}\BibitemShut {NoStop}%
\bibitem [{\citenamefont {{Manzotti}}\ and\ \citenamefont
  {{Dietz}}(2012)}]{2012arXiv1202.4031M}%
  \BibitemOpen
  \bibfield  {author} {\bibinfo {author} {\bibfnamefont {A.}~\bibnamefont
  {{Manzotti}}}\ and\ \bibinfo {author} {\bibfnamefont {A.}~\bibnamefont
  {{Dietz}}},\ }\href@noop {} {\bibfield  {journal} {\bibinfo  {journal} {arXiv
  e-prints}\ } (\bibinfo {year} {2012})},\ \Eprint
  {http://arxiv.org/abs/1202.4031} {arXiv:1202.4031 [gr-qc]} \BibitemShut
  {NoStop}%
\bibitem [{\citenamefont {Seto}(2006)}]{Seto:2005qy}%
  \BibitemOpen
  \bibfield  {author} {\bibinfo {author} {\bibfnamefont {N.}~\bibnamefont
  {Seto}},\ }\href {\doibase 10.1103/PhysRevD.73.063001} {\bibfield  {journal}
  {\bibinfo  {journal} {Phys.Rev.}\ }\textbf {\bibinfo {volume} {D73}},\
  \bibinfo {pages} {063001} (\bibinfo {year} {2006})},\ \Eprint
  {http://arxiv.org/abs/gr-qc/0510067} {arXiv:gr-qc/0510067 [gr-qc]}
  \BibitemShut {NoStop}%
\bibitem [{\citenamefont {{Kuroyanagi}}\ \emph {et~al.}(2018)\citenamefont
  {{Kuroyanagi}}, \citenamefont {{Chiba}},\ and\ \citenamefont
  {{Takahashi}}}]{Kuroyanagi_2018JCAP}%
  \BibitemOpen
  \bibfield  {author} {\bibinfo {author} {\bibfnamefont {S.}~\bibnamefont
  {{Kuroyanagi}}}, \bibinfo {author} {\bibfnamefont {T.}~\bibnamefont
  {{Chiba}}}, \ and\ \bibinfo {author} {\bibfnamefont {T.}~\bibnamefont
  {{Takahashi}}},\ }\href {\doibase 10.1088/1475-7516/2018/11/038} {\bibfield
  {journal} {\bibinfo  {journal} {\jcap}\ }\textbf {\bibinfo {volume} {2018}},\
  \bibinfo {eid} {038} (\bibinfo {year} {2018})},\ \Eprint
  {http://arxiv.org/abs/1807.00786} {arXiv:1807.00786 [astro-ph.CO]}
  \BibitemShut {NoStop}%
\bibitem [{\citenamefont {Himemoto}\ \emph {et~al.}(2021)\citenamefont
  {Himemoto}, \citenamefont {Nishizawa},\ and\ \citenamefont
  {Taruya}}]{Himemoto:2021ukb}%
  \BibitemOpen
  \bibfield  {author} {\bibinfo {author} {\bibfnamefont {Y.}~\bibnamefont
  {Himemoto}}, \bibinfo {author} {\bibfnamefont {A.}~\bibnamefont {Nishizawa}},
  \ and\ \bibinfo {author} {\bibfnamefont {A.}~\bibnamefont {Taruya}},\ }\href
  {\doibase 10.1103/PhysRevD.104.044010} {\bibfield  {journal} {\bibinfo
  {journal} {Phys. Rev. D}\ }\textbf {\bibinfo {volume} {104}},\ \bibinfo
  {pages} {044010} (\bibinfo {year} {2021})},\ \Eprint
  {http://arxiv.org/abs/2103.14816} {arXiv:2103.14816 [gr-qc]} \BibitemShut
  {NoStop}%
\bibitem [{\citenamefont {Abbott}\ \emph {et~al.}(2017)\citenamefont {Abbott}
  \emph {et~al.}}]{Evans:2016mbw}%
  \BibitemOpen
  \bibfield  {author} {\bibinfo {author} {\bibfnamefont {B.~P.}\ \bibnamefont
  {Abbott}} \emph {et~al.} (\bibinfo {collaboration} {LIGO Scientific}),\
  }\href {\doibase 10.1088/1361-6382/aa51f4} {\bibfield  {journal} {\bibinfo
  {journal} {Class. Quant. Grav.}\ }\textbf {\bibinfo {volume} {34}},\ \bibinfo
  {pages} {044001} (\bibinfo {year} {2017})},\ \Eprint
  {http://arxiv.org/abs/1607.08697} {arXiv:1607.08697 [astro-ph.IM]}
  \BibitemShut {NoStop}%
\bibitem [{ET:()}]{ET:2020}%
  \BibitemOpen
  \href@noop {} {}\bibinfo {note} {Design Report Update 2020 for the Einstein
  Telescope: https://apps.et-gw.eu/tds/?content=3\&r=17245}\BibitemShut
  {NoStop}%
\bibitem [{\citenamefont {{LIGO}}\ \emph {et~al.}(2022)\citenamefont {{LIGO}},
  \citenamefont {{Virgo}},\ and\ \citenamefont {{KAGRA
  collaborations}}}]{LVK:2022}%
  \BibitemOpen
  \bibfield  {author} {\bibinfo {author} {\bibnamefont {{LIGO}}}, \bibinfo
  {author} {\bibnamefont {{Virgo}}}, \ and\ \bibinfo {author} {\bibnamefont
  {{KAGRA collaborations}}},\ }\href
  {https://dcc.ligo.org/LIGO-T2000012/public} {\bibfield  {journal} {\bibinfo
  {journal} {https://dcc.ligo.org/LIGO-T2000012/public}\ } (\bibinfo {year}
  {2022})}\BibitemShut {NoStop}%
\end{thebibliography}
\bibliographystyle{apsrev4-1}
%


\end{document}